\documentclass[onecolumn,nofootinbib,superscriptaddress,fleqn,amssymb,amsfonts,amsmath]{revtex4} 
\usepackage{graphicx,slashed,xcolor,multirow,bbold,mathtools,sidecap,tikz,bm,ulem,enumitem,booktabs,array}
\usepackage[colorlinks=true,linktoc=page,linkcolor=purple,citecolor=teal,urlcolor=magenta]{hyperref}
\usepackage[compat=1.1.0]{tikz-feynman}

\definecolor{lime}{HTML}{A6CE39}
\DeclareRobustCommand{\orcidicon}{\hspace{-2.1mm}
\begin{tikzpicture}
\draw[lime,fill=lime] (0,0.0) circle [radius=0.13] node[white] {{\fontfamily{qag}\selectfont \tiny ID}}; \draw[white,fill=white] (-0.0525,0.095) circle [radius=0.007];
\end{tikzpicture} \hspace{-3.7mm} }
\foreach \x in {A, ..., Z}
 {\expandafter\xdef\csname orcid\x\endcsname{\noexpand\href{https://orcid.org/\csname orcidauthor\x\endcsname} {\noexpand\orcidicon}}}

\begin{document}

\title{More constraints on the Georgi-Machacek model}

\author{Zahra Bairi\orcidA{}}
\email{z.bairi@univ-boumerdes.dz}
\affiliation{Laboratory of Photonic Physics and Nano-Materials, Department of Matter
Sciences, University of Biskra, DZ-07000 Biskra, Algeria}
\affiliation{Laboratoire de Physique des Particules et Physique Statistique, Ecole
Normale Superieure, BP 92 Vieux Kouba, DZ-16050 Algiers, Algeria}
\affiliation{Department of Physics, University of M'Hamed Bougara-Boumerdes, DZ-35000
Boumerdes, Algeria.}

\author{Amine Ahriche\orcidB{}}
\email{ahriche@sharjah.ac.ae}
\affiliation{Department of Applied Physics and Astronomy, University of Sharjah,
P.O. Box 27272 Sharjah, United Arab Emirates}
\affiliation{The Abdus Salam International Centre for Theoretical Physics, Strada
Costiera 11, I-34014, Trieste, Italy}
\affiliation{Laboratoire de Physique des Particules et Physique Statistique, Ecole
Normale Superieure, BP 92 Vieux Kouba, DZ-16050 Algiers, Algeria}

\begin{abstract}
In this work, we investigate the parameter space of the Georgi-Machacek
(GM) model, where we consider many theoretical and experimental constraints
such as the perturbativity, vacuum stability, unitarity, electroweak
precision tests, the Higgs diphoton decay, the Higgs total decay
width and the LHC measurements of the signal strengths of the SM-like
Higgs boson $h$ in addition to the constraints from doubly charged
Higgs bosons and Drell-Yan diphoton production and the indirect constraint
from the $b\to s$ transition processes. We investigate also
the possibility that the electroweak vacuum could be destabilized
by unwanted wrong minima that may violate the $CP$ and/or the electric
charge symmetries. We found that about 40 \% of the parameter space
that fulfills the above mentioned constraints are excluded by these
unwanted minima. In addition, we found that the negative searches
for a heavy resonance could exclude a significant part of the viable
parameter space, and future searches could exclude more regions in
the parameter space.
\end{abstract}

\maketitle

\section{Introduction}

Since the discovery of a Standard Model (SM)-like 125 GeV Higgs boson
at the Large Hadron Collider (LHC)~\cite{ATLAS:2012yve}, many questions
are still open, where the SM provides no answers. For instance, the
Higgs mass is found to be at the electroweak (EW) scale, while it
may acquire very large radiative corrections that can reach the Planck
or GUT scales within the SM. This hierarchy problem requires an unwanted
fine-tuning. In addition, there are unanswered questions such as the
fermions masses of difference, the origin of $CP$ violation in the quark
sector, the dark matter nature~\cite{Bertone:2004pz} and the neutrino
oscillation data~\cite{Super-Kamiokande:1998kpq}.

The discovered 125 GeV scalar has the properties of a SM-like Higgs;
however, it is not known yet whether the electroweak symmetry breaking
(EWSB) is triggered by one single scalar field or more. In many SM
extensions, the EWSB is achieved via more than one scalar where many
scalar fields acquire nonvanishing vacuum expectation values (VEVs),
and the SM-like is a composite. Among these SM extensions, the so-called
Georgi-Machacek (GM) model~\cite{Georgi:1985nv}, where the ESWB
is realized by three scalar fields. In addition to the SM doublet, the
GM model includes one complex and one real scalar triplets, where
a global custodial $SU(2)_V$ symmetry is preserved in the scalar
potential after the EWSB. The scalar vacuum in the GM model is defined
in a way that the $\rho$-parameter should be within the experimentally
allowed range~\cite{Workman:2022ynf},
\begin{equation}
\rho=\frac{g_{hWW}^{SM}}{g_{hZZ}^{SM}\cos^{2}\theta_{w}}=1.00039\pm0.00019,\label{eq:rho}
\end{equation}
with $g_{hWW}^{SM}=2m_{W}^{2}/\upsilon$ and $g_{hZZ}^{SM}=2m_{Z}^{2}/\upsilon$,
where $\upsilon=246.22\,\textrm{GeV}$. This leads to a scalar spectrum
with different multiplets under the global $SU(2)_V$ custodial
symmetry, whose mass eigenstates give a quintet ($H_{5}$), a triplet
($H_{3}$) and two $CP-even$ singlets ($\eta$ and $h$). In our work,
we consider the parameter space that corresponds to $h=h_{125}$,
with $m_{\eta}>m_{h}$. One has to mention that an interesting viable
parameter space exists for the case $m_{\eta}<m_{h}$, where interesting
collider signatures are possible~\cite{Ahriche:2022aoj}.

Due to the feature that the SM-like Higgs couplings to both $W$ and
$Z$ gauge bosons could be significantly different than the SM values~\cite{Chiang:2013rua},
the GM model could be phenomenologically interesting. In addition
to the existence of additional $CP-odd$, singly and doubly charged scalars,
the GM model could be a good benchmark for searches of beyond SM scalars;
which has been extensively investigated in the literature~\cite{Chanowitz:1985ug}.
In the decoupling limit~\cite{Hartling:2014zca}, all additional
beyond SM particles that are present in the GM model become heavy
and the fermion and gauge bosons couplings to the SM-like Higgs boson
approach the SM values. In addition to the rich phenomenology, other
issues were addressed within the GM model such as the neutrino mass~\cite{Chen:2020ark},
dark matter~\cite{Pilkington:2017qam}, and the electroweak phase
transition strength~\cite{Chiang:2014hia}.

Recent measurements and negative searches at the LHC~\cite{Workman:2022ynf},
such as those of the total decay width, Higgs strength modifiers and
the cross section upper bounds from negative searches of new scalar
resonance, could imply significant constraints on the GM model parameter
space. Although the GM model includes a custodial scalar fiveplet, it has been shown that the LHC searches for the doubly charged
Higgs bosons in the VBF channel $H_{5}^{++}\to W^{+}W^{+}$
and the Drell-Yan production of a neutral Higgs boson $pp\to H_{5}^{0}(\gamma\gamma)H^{+}$
impose interesting bounds on the parameter space~\cite{Ismail:2020zoz},
as well as the indirect constraints from the $b\to s$ transition
processes that exclude all the benchmark points (BPs) with large $\upsilon_{\xi}$~\cite{Hartling:2014aga}.
In addition, the GM scalar potential structure may admit many minima
beside the electroweak (EW) vacuum that could break the electric
charge and/or the $CP$ symmetry spontaneously. In case where such minima
exist, they should not be deeper than the EW vacuum, which may affect
the parameter space that is in agreement with the previously mentioned
constraints. In~\cite{Chiang:2018cgb}, the authors performed a global
fit analysis for the GM model free parameter and obtained some limits
on the mixing angles and the heavy new scalar masses and decay widths.
However, since the constraints from the $b\to s$ transition
processes were considered in~\cite{Chiang:2018cgb}; and the LHC
measurements used to constrain the GM model have been significantly
updated, an analysis for the full model parameter space is required.
Here, we aim to investigate the impact of all the relevant constraints
on the model by performing a full numerical scan over the whole parameter
space.

In this work, we give a brief introduction of the GM model in Sec~\ref{sec:Model},
where the scalar potential and the mass spectrum are described. In
Sec~\ref{sec:WrM}, we discuss the possible existence of new
minima that could be deeper than the EW vacuum. Then, after categorizing
these unwanted minima according to the preserved/broken ($CP$ and electric
charge) symmetries, one considers the EW vacuum to be the deepest one
as an novel constraint on the GM model. In Sec~\ref{sec:Constraints},
we discuss different theoretical and experimental constraints on the
model such as the unitarity, vacuum stability, the total Higgs decay
width and signal strength modifiers, the electroweak precision tests,
and the diphoton Higgs decay. In addition, we consider the recent
ATLAS and CMS constraints on the heavy $CP-even$ scalar $\eta$ and
from the negative searches for the doubly charged Higgs bosons in
the VBF channel $H_{5}^{++}\to W^{+}W^{+}$, and the Drell-Yan
production of a neutral Higgs boson $pp\to H_{5}^{0}(\gamma\gamma)H^{+}$.
We show our numerical results and discussion in Sec~\ref{sec:NA},
and our conclusion in Sec~\ref{sec:Conclusion}.

\section{The Model: Parameters and Mass Spectrum\label{sec:Model}}

In the GM model, the scalar sector consists of a scalar doublet $(\phi^{+},\,\phi^{0})^{T}$
with hypercharge $Y=1$, and two triplet representations $(\chi^{++},\,\chi^{+},\,\chi^{0})^{T}$
and $(\xi^{+},\,\xi^{0},\,-\xi^{-})^{T}$ with hypercharge $Y=2,0$,
respectively. These representations can be written as
\begin{equation}
\Phi=\left(\begin{array}{cc}
\phi^{0*} & \phi^{+}\\
-\phi^{+*} & \phi^{0}
\end{array}\right),\quad\Delta=\left(\begin{array}{ccc}
\chi^{0*} & \xi^{+} & \chi^{++}\\
-\chi^{+*} & \xi^{0} & \chi^{+}\\
\chi^{++*} & -\xi^{+*} & \chi^{0}
\end{array}\right),\label{eq:field}
\end{equation}
where $\phi^{-}=\phi^{+*},\,\xi^{-}=\xi^{+*},\,\chi^{--}=\chi^{++*},\,\chi^{-}=\chi^{+*}.$
The neutral components in (\ref{eq:field}) can be expressed by
\begin{equation}
\phi^{0}=\frac{1}{\sqrt{2}}(\upsilon_{\phi}+h_{\phi}+ia_{\phi}),\,\chi^{0}=\frac{1}{\sqrt{2}}(\upsilon_{\chi}+h_{\chi}+ia_{\chi}),\,\xi^{0}=\upsilon_{\xi}+h_{\xi},\label{eq:vevs}
\end{equation}
where $\upsilon_{\phi},\,\upsilon_{\chi}$ and $\upsilon_{\xi}$ are
the VEVs for $\phi^{0},\,\chi^{0}$ and $\xi^{0}$, respectively.
Here, we have three $CP-even$ scalar degrees of freedom (d.o.f.) $\{h_{\phi},\,h_{\chi},\,h_{\xi}\}$,
two $CP-odd$ d.o.f. $\{a_{\phi},\,a_{\chi}\}$, six singly charged d.o.f.
$\{\phi^{\pm},\chi^{\pm},\xi^{\pm}\}$ and two doubly charged d.o.f.
$\chi^{\pm\pm}$. The most general scalar potential invariant under
the global symmetry $SU(2)_{L}\times SU(2)_{R}\times U(1)_{Y}$ is
given by
\begin{align}
V(\varPhi,\Delta) & =\frac{m_{1}^{2}}{2}\mathrm{Tr}[\varPhi^{\dagger}\varPhi]+\frac{m_{2}^{2}}{2}\mathrm{Tr}[\Delta^{\dagger}\Delta]+\lambda_{1}(\mathrm{Tr}[\varPhi^{\dagger}\varPhi])^{2}+\lambda_{2}\mathrm{Tr}[\varPhi^{\dagger}\varPhi]\mathrm{Tr}[\Delta^{\dagger}\Delta]\nonumber \\
 & +\lambda_{3}\mathrm{Tr}[(\Delta^{\dagger}\Delta)^{2}]+\lambda_{4}(\mathrm{Tr}[\Delta^{\dagger}\Delta])^{2}-\lambda_{5}\mathrm{Tr}[\varPhi^{\dagger}\frac{\sigma^{a}}{2}\varPhi\frac{\sigma^{b}}{2}]\mathrm{Tr}[\Delta^{\dagger}T^{a}\Delta T^{b}]\nonumber \\
 & -\mu_{1}\mathrm{Tr}[\varPhi^{\dagger}\frac{\sigma^{a}}{2}\varPhi\frac{\sigma^{b}}{2}](U\Delta U^{\dagger})_{ab}-\mu_{2}\mathrm{Tr}[\Delta^{\dagger}T^{a}\Delta T^{b}](U\Delta U^{\dagger})_{ab},\label{eq:V}
\end{align}
with $\sigma^{1,2,3}$ are the Pauli matrices and $T^{1,2,3}$ correspond
to the generators of the $SU(2)$ triplet representation, that are
given by
\begin{equation}
T^{1}=\frac{1}{\sqrt{2}}\left(\begin{array}{ccc}
0 & 1 & 0\\
1 & 0 & 1\\
0 & 1 & 0
\end{array}\right),\:T^{2}=\frac{1}{\sqrt{2}}\left(\begin{array}{ccc}
0 & -i & 0\\
i & 0 & -i\\
0 & i & 0
\end{array}\right),\,T^{3}=\left(\begin{array}{ccc}
1 & 0 & 0\\
0 & 0 & 0\\
0 & 0 & -1
\end{array}\right),
\end{equation}
and the matrix $U$ is defined as
\begin{equation}
U=\frac{1}{\sqrt{2}}\left(\begin{array}{ccc}
-1 & 0 & 1\\
-i & 0 & -i\\
0 & \sqrt{2} & 0
\end{array}\right).
\end{equation}

The custodial symmetry condition at tree-level $m_{W}^{2}=m_{Z}^{2}\cos^{2}\theta_{W}$
implies $\upsilon_{\chi}=\sqrt{2}\upsilon_{\xi}$ and $\upsilon_{\phi}^{2}+8\upsilon_{\xi}^{2}\equiv\upsilon^{2}=(246.22\,\mathrm{GeV})^{2}$,
where $m_{W}$, $m_{Z}$ and $\theta_{W}$ are the gauge bosons masses
and the Weinberg mixing angle. It would be useful to introduce the
parameter $t_{\beta}\equiv\tan\beta=2\sqrt{2}\upsilon_{\xi}/\upsilon_{\phi}$
to describe the relations between the VEV's. By using the tadpole
conditions, one can eliminate the parameters $m_{1,2}^{2}$ as
\begin{align}
m_{1}^{2} & =-4\lambda_{1}c_{\beta}^{2}\upsilon^{2}+\frac{3}{8}(-2\lambda_{2}+\lambda_{5})s_{\beta}^{2}\upsilon^{2}+\frac{3}{4\sqrt{2}}\mu_{1}s_{\beta}\upsilon,\nonumber \\
m_{2}^{2} & {\it =}(-2\lambda_{2}+\lambda_{5})c_{\beta}^{2}\upsilon^{2}-\frac{1}{2}(\lambda_{3}+3\,\lambda_{4})s_{\beta}^{2}\upsilon^{2}+\frac{\mu_{1}}{\sqrt{2}}\frac{c_{\beta}^{2}\upsilon}{s_{\beta}}+\frac{3}{\sqrt{2}}\mu_{2}s_{\beta}\upsilon.
\end{align}

After the EWSB, the Goldstone bosons are eaten by the massive W and
Z bosons, and we are left with the following mass eigenstates: three
$CP-even$ eigenstates $\{h,\eta,H_{5}^{0}\}$, one $CP-odd$ eigenstate
$H_{3}^{0}$, two singly charged scalars $\{H_{3}^{\pm},H_{5}^{\pm}\}$,
and one doubly charged scalar $H_{5}^{\pm\pm}$,
\begin{align}
h & =c_{\alpha}h_{\phi}-\frac{s_{\alpha}}{\sqrt{3}}(\sqrt{2}h_{\chi}+h_{\xi}),\,\eta=s_{\alpha}h_{\phi}+\frac{c_{\alpha}}{\sqrt{3}}(\sqrt{2}h_{\chi}+h_{\xi}),\,H_{5}^{0}=\sqrt{\frac{2}{3}}h_{\xi}-\sqrt{\frac{1}{3}}h_{\chi},\nonumber \\
H_{3}^{0} & =-s_{\beta}a_{\phi}+c_{\beta}a_{\chi},\,H_{3}^{\pm}=-s_{\beta}\phi^{\pm}+c_{\beta}\frac{1}{\sqrt{2}}(\chi^{\pm}+\xi^{\pm}),\,H_{5}^{\pm}=\frac{1}{\sqrt{2}}(\chi^{\pm}-\xi^{\pm}),\,H_{5}^{\pm\pm}=\chi^{\pm\pm}.\label{eq:Eigen}
\end{align}

The mixing angle $\alpha$ of the $CP-even$ sector can defined by $\tan2\alpha=2M_{12}^{2}/(M_{22}^{2}-M_{11}^{2})$,
where $M^{2}$ is the mass squared matrix in the basis $\{h_{\phi},\,\sqrt{\frac{2}{3}}h_{\chi}+\frac{1}{\sqrt{3}}h_{\xi}\}$,
whose elements are given by
\begin{align}
M_{11}^{2} & =8\,\lambda_{1}c_{\beta}^{2}\upsilon^{2},\nonumber \\
M_{12}^{2} & =\frac{\sqrt{3}}{2}c_{\beta}\upsilon[-\mu_{1}+\sqrt{2}(2\,\lambda_{2}-\lambda_{5})s_{\beta}\upsilon],\nonumber \\
M_{22}^{2} & =\frac{\mu_{1}}{\sqrt{2}}\frac{c_{\beta}^{2}\upsilon}{s_{\beta}}-\frac{3}{\sqrt{2}}\,\mu_{2}s_{\beta}\upsilon+(\lambda_{3}+3\,\lambda_{4})s_{\beta}^{2}\upsilon^{2}.
\end{align}

This allows us to write the SM-like Higgs bosons and the heavy scalar ($\eta$)
eigenmasses as $m_{h,\eta}^{2}=\frac{1}{2}[M_{11}^{2}+M_{22}^{2}\mp\sqrt{(M_{11}^{2}-M_{22}^{2})^{2}+4(M_{12}^{2})^{2}}]$.
The other eigenmasses are
\begin{align}
m_{H_{3}^{0}}^{2} & =m_{H_{3}^{\pm}}^{2}=m_{3}^{2}=(\frac{\mu_{1}}{\sqrt{2}s_{\beta}\upsilon}+\frac{\lambda_{5}}{2})\,\upsilon^{2},\nonumber \\
m_{H_{5}^{0}}^{2} & =m_{H_{5}^{\pm}}^{2}=m_{H_{5}^{\pm\pm}}^{2}=m_{5}^{2}=\frac{\mu_{1}}{\sqrt{2}}\frac{c_{\beta}^{2}\upsilon}{s_{\beta}}+\frac{6}{\sqrt{2}}\,\mu_{2}s_{\beta}\upsilon+\frac{3}{2}\,\lambda_{5}c_{\beta}^{2}\upsilon^{2}+\lambda_{3}s_{\beta}^{2}\upsilon^{2}.
\end{align}

Since, we will take the masses as input parameters, the quartic couplings
$\lambda$'s can be expressed as
\begin{align}
\lambda_{1} & =\frac{\varrho_1 c_{\alpha}^{2}+\varrho_2 s_{\alpha}^{2}}{8\upsilon^{2}c_{\beta}^{2}},\,\lambda_{2}=-\frac{c_{\alpha}s_{\alpha}(\varrho_{1}-\varrho_{2})}{\sqrt{6}\upsilon^{2}c_{\beta}s_{\beta}}+\frac{m_{3}^{2}}{\upsilon^{2}}-\frac{\mu_{1}}{2\sqrt{2}\upsilon s_{\beta}},\nonumber \\
\lambda_{3} & =-\frac{3c_{\beta}^{2}m_{3}^{2}}{s_{\beta}^{2}\upsilon^{2}}+\frac{m_{5}^{2}}{s_{\beta}^{2}\upsilon^{2}}+\frac{\sqrt{2}(\mu_{1}c_{\beta}^{2}-3\mu_{2}s_{\beta}^{2})}{s_{\beta}^{3}\upsilon},\,\lambda_{5}=\frac{2m_{3}^{2}}{\upsilon^{2}}-\frac{\sqrt{2}\mu_{1}}{\upsilon s_{\beta}},\nonumber \\
\lambda_{4} & =\frac{\varrho_1 s_{\alpha}^{2}+\varrho_2 c_{\alpha}^{2}}{3s_{\beta}^{2}\upsilon^{2}}+\frac{c_{\beta}^{2}m_{3}^{2}}{s_{\beta}^{2}\upsilon^{2}}-\frac{m_{5}^{2}}{3s_{\beta}^{2}\upsilon^{2}}-\frac{\mu_{1}c_{\beta}^{2}-3\mu_{2}s_{\beta}^{2}}{\sqrt{2}s_{\beta}^{3}\upsilon}
,\label{eq:lm}
\end{align}
with $\varrho_{1}=\min(m_{h}^2,m_{\eta}^2)$ and $\varrho_{2}=\max(m_{h}^2,m_{\eta}^2)$.
The formulas of $\lambda_{1,2,4}$ here are valid for both cases of
$m_{h}<m_{\eta}$ and $m_{h}>m_{\eta}$.

\section{Avoiding wrong minima\label{sec:WrM}}

Since the scalar potential is a function of different fields; three $CP-even$,
two $CP-odd$ and eight charged scalars, the possibility of other existing
minima that are different and deeper than $(\Re(\phi^{0}),\Re(\chi^{0}),\Re(\xi^{0}))=(\upsilon_{\phi},\,\sqrt{2}\upsilon_{\xi},\upsilon_{\xi})$
would destabilize the EW vacuum. In~\cite{Hartling:2014zca,Moultaka:2020dmb},
the authors adopted a simplified field parametrization to investigate
the vacuum stability and the boundness from below conditions, where
the scalar potential (\ref{eq:V}) can be written as
\begin{align}
V & =\frac{1}{2}\frac{r^{2}}{(1+\tan^{2}\gamma)}[m_{1}^{2}+m_{2}^{2}\tan^{2}\gamma]+\frac{r^{3}}{(1+\tan^{2}\gamma)^{3/2}}\tan\gamma[-\sigma\mu_{1}-\rho\mu_{2}\tan^{2}\gamma]\nonumber \\
 & +\frac{r^{4}}{(1+\tan^{2}\gamma)^{2}}[\lambda_{1}+(\lambda_{2}-\omega\lambda_{5})\tan^{2}\gamma+(\zeta\lambda_{3}+\lambda_{4})\tan^{4}\gamma],\label{eq:Vp}
\end{align}
with
\begin{align}
r & =\sqrt{\mathrm{Tr}(\varPhi^{\dagger}\varPhi)+\mathrm{Tr}(\Delta^{\dagger}\Delta)},\,\mathrm{Tr}(\varPhi^{\dagger}\varPhi)=r^{2}\cos^{2}\gamma,\,\mathrm{Tr}(\Delta^{\dagger}\Delta)=r^{2}\sin^{2}\gamma,\,\nonumber \\
 & \mathrm{Tr}(\Delta^{\dagger}\Delta\Delta^{\dagger}\Delta)=\zeta\,r^{4}\sin^{4}\gamma,\,\mathrm{Tr}(\varPhi^{\dagger}\sigma^{a}\varPhi\sigma^{b})\mathrm{Tr}(\Delta^{\dagger}T^{a}\Delta T^{b})=\omega r^{4}\cos^{2}\gamma,\,\sin^{2}\gamma,\nonumber \\
 & \mathrm{Tr}(\varPhi^{\dagger}\sigma^{a}\varPhi\sigma^{b})(U\Delta U^{\dagger})_{ab}=\sigma r^{3}\sin\gamma\cos^{2}\gamma,\ \mathrm{Tr}(\Delta^{\dagger}T^{a}\Delta T^{b})(U\Delta U^{\dagger})_{ab}=\rho r^{3}\sin^{3}\gamma,\nonumber \\
 & r\in[1,\infty[,\gamma\in[0,\frac{\pi}{2}],\ \zeta\in[\frac{1}{3},1],\omega\in[-\frac{1}{4},\frac{1}{2}],\,\sigma\in[-\frac{\sqrt{3}}{4},\frac{\sqrt{3}}{4}],\ \rho\in[-\frac{2}{\sqrt{3}},\frac{2}{\sqrt{3}}].\label{eq:parameters}
\end{align}

For instance, the conditions for the boundness from below of the scalar
potential can be ensured by imposing the coefficients of the quartic
term i.e., the second line in (\ref{eq:Vp}) to be positive, which
leads to
\begin{equation}
\lambda_{1}>0,\ \zeta\lambda_{3}+\lambda_{4}>0,\ \lambda_{2}-\omega\lambda_{5}+2\sqrt{\lambda_{1}(\zeta\lambda_{3}+\lambda_{4})}>0.
\end{equation}

The parametrization (\ref{eq:Vp}) reduces the searches for the potential
minima into looking for specific sets of the parameters values in
the ranges (\ref{eq:parameters}) that make (\ref{eq:Vp}) minimal.
Here, we will not adopt this approach due to many reasons, among them
the fact that the parameters in (\ref{eq:parameters}) are not fully
independent. In other words, any field configuration in the field
space can be defined by a single set of the parameters in (\ref{eq:parameters}),
while any parameters set in (\ref{eq:parameters}) does not necessarily
correspond to a well-defined field configuration. In addition, when
a field configuration corresponds to a minimum, it does not show whether
it preserves or violates the $CP$ symmetry and/or the electric charge.

The scalar potential includes 13 scalar d.o.f.: three $CP-even$, two $CP-odd$,
six singly charged and two doubly charged. The scalar potential must respect
the electric charge conservation by demanding (1) either the VEVs
of all charged scalars to be vanishing, i.e., $<\phi^{\pm}>=<\chi^{\pm}>=<\xi^{\pm}>=<\chi^{++}>=0$,
or (2) any existing electric charge breaking minimum should not be
deeper than the EW one. The $CP$ symmetry could be spontaneously violated
when some of the $CP-odd$ fields acquire a VEV, i.e., $<\Im(\phi^{0})>,<\Im(\chi^{0})>\neq0$,
where this case is experimentally allowed within the data from ACME
Collaboration on the electron and neutron electric dipole moment (EDM)~\cite{ACME:2013pal}.
In the case where both $CP$ symmetry and the electric charge are conserved,
other minima beside the EW vacuum $(\Re(\phi^{0}),\Re(\chi^{0}),\Re(\xi^{0}))=(\upsilon_{\phi},\,\sqrt{2}\upsilon_{\xi},\upsilon_{\xi})$,
could exist. In order to ensure the EW vacuum stability, we need to
check that the scalar potential at $(\Re(\phi^{0}),\Re(\chi^{0}),\Re(\xi^{0}))=(\upsilon_{\phi},\,\sqrt{2}\upsilon_{\xi},\upsilon_{\xi})$
is the true global minimum. Then, in our work we consider only the
parameter space where the EW vacuum is deeper than an any other existing
minimum whether it preserves or violates the $CP$ and/or electric charge
symmetries.

Then, finding these wrong minima requires the minimization of the
potential (\ref{eq:V}) along all the $CP-even$, $CP-odd$ and the charged
fields directions is mandatory. As the minimization along the $CP-odd$
2D space $\{\Im(\phi^{0}),\Im(\chi^{0})\}$ is straightforward, it
requires along the charged directions a useful parametrization for
the charged fields. This can be done either by writing both singly
and doubly charged fields as $X^{\pm}=\frac{1}{\sqrt{2}}(x_{1}\pm i x_{2})$~\cite{Azevedo:2020mjg}, or adopting the parametrization $X^{\pm}=|X|e^{\pm i\varrho}$. In~\cite{Azevedo:2020mjg}, the authors studied the vacuum stability of a $Z_2$ symmetric version of the GM model, where the cubic terms of the scalar potential are absent. They used the parametrization $X^{\pm}=\frac{1}{\sqrt{2}}(x_{1}\pm i x_{2})$ to investigate special cases in which $CP$ and/or electric charge symmetries could be violated. However, this study is not applicable to our research due to the global $Z_2$ symmetry (i.e., $\mu_1=\mu_2=0$), which renders the possible vacua drastically different from the standard case where $\mu_1$ and $\mu_2$ are nonzero.

In our work, we consider the polar parametrization where the minimization
conditions are $\partial V/\partial X=\partial V/\partial\varrho=0$
at the charge breaking vacuum. Although in the $CP-even$ directions,
there may exist other minima beside the EW one that could be deeper.
Therefore, one has to search for all minima along all directions ($CP-even$,
$CP-odd$ and charged) and check that they are not deeper than the EW
vacuum $(\upsilon_{\phi},\,\sqrt{2}\upsilon_{\xi},\upsilon_{\xi})$.

After a careful analysis, we found eight minima in the $CP-even$ directions
$\{h_{\phi},\,h_{\chi},\,h_{\xi}\}$, three minima along the $CP-odd$ directions
$\{a_{\phi},\,a_{\chi}\}$, eight minima along the singlet charged fields
directions $\{\phi^{\pm},\chi^{\pm},\xi^{\pm}\}$, and a minimum along
the doubly charged direction $\chi^{\pm\pm}$. We denote the potential
values at these wrong minima by $V_{i=1,8}^{0+}$, $V_{i=1,3}^{0-},$
$V_{i=1,8}^{\pm}$ and $V^{\pm}$, respectively, and we give their
coordinates in {Appendix~\ref{app:WM}. Getting the analytical
formula for the $CP$-conserving and electric charge violating minima
given in (\ref{eq:C1}), (\ref{eq:C2}) and (\ref{eq:C3}) was an
easy task since they were special cases of one or two-dimensional
problem. Indeed, there could be other minima defined in 3D, which
will be defined numerically.

Then, the EW vacuum should be deeper than all these local minima,
i.e.,
\begin{equation}
V(\Re(\phi^{0})=\upsilon_{\phi},\Re(\chi^{0})=\sqrt{2}\upsilon_{\xi},\Re(\xi^{0})\,=\upsilon_{\xi})<\min\big\{ V_{i}^{0+},V_{i}^{0-},V_{i}^{\pm},V^{\pm\pm},0\big\},\label{eq:Vac}
\end{equation}
where the zero in the last position represents the obviously wrong
vacuum $V(0,0,0)$. As we will see later, the condition (\ref{eq:Vac})
could exclude more than 40 \% of the parameter space.

\section{Theoretical and Experimental Constraints\label{sec:Constraints}}

In what follows, we discuss different theoretical and experimental
constraints on the GM model that are related to many aspects such
as the vacuum stability, unitarity, the Higgs decays, the electroweak
precision tests, in addition to the constraints from negative searches
for heavy scalar resonances at the LHC.

\vspace{0.5cm}

\textbf{Tree-level unitarity}

The bound from perturbative unitarity is obtained by requiring the
zeroth partial wave amplitude for any elastic $2\to 2$ bosonic
scatterings does not become too large to violate $S$ matrix unitarity.
In the high $CM$ energy regime, the gauge fields can be replaced by
their corresponding Goldstone scalars. This means that the amplitude,
$a_{0}$ satisfy $\left|a_{0}\right|\leq1$ or $\left|Re\,a_{0}\right|\leq1/2$.
Then, the perturbative unitarity bounds in the GM model reads~\cite{Hartling:2014zca}\textbf{
\begin{align}
\sqrt{(6\lambda_{1}-7\lambda_{3}-11\lambda_{4})^{2}+36\lambda_{2}^{2}}+\left|6\lambda_{1}+7\lambda_{3}+11\lambda_{4}\right| & <4\pi,\,\left|2\lambda_{3}+\lambda_{4}\right|<\pi,\nonumber \\
\sqrt{(2\lambda_{1}+\lambda_{3}-2\lambda_{4})^{2}+\lambda_{5}^{2}}+\left|2\lambda_{1}-\lambda_{3}+2\lambda_{4}\right|<4\pi, & \,\,\left|\lambda_{2}-\lambda_{5}\right|<2\pi.
\end{align}
}

\vspace{0.5cm}

\textbf{Boundness from below}

To ensure the scalar potential boundness from below condition, the
coefficients of the quartic term along any direction in the fields
space must be positive. This leads to the conditions~\cite{Arhrib:2011uy}

\begin{align}
\lambda_{1} & >0,\,\lambda_{4}>\left\{ \begin{array}{c}
-\frac{1}{3}\lambda_{3}\quad\textrm{for}\ \lambda_{3}\geq0,\\
-\lambda_{3}\qquad\textrm{for}\quad\lambda_{3}<0,
\end{array}\right.,\nonumber \\
\lambda_{2} & >\begin{cases}
\begin{array}{c}
\frac{1}{2}\lambda_{5}-2\sqrt{\lambda_{1}\left(\frac{1}{3}\lambda_{3}+\lambda_{4}\right)}\hfill\mathrm{\textrm{for}}\ \lambda_{5}\geq0\ \textrm{and}\ \lambda_{3}\geq0,\\
\omega_{+}(\zeta)\lambda_{5}-2\sqrt{\lambda_{1}\left(\zeta\lambda_{3}+\lambda_{4}\right)}\qquad\textrm{for}\ \lambda_{5}\geq0\ \textrm{and}\ \lambda_{3}<0,\\
\omega_{-}(\zeta)\lambda_{5}-2\sqrt{\lambda_{1}\left(\zeta\lambda_{3}+\lambda_{4}\right)}~~~\textrm{ for}\ \lambda_{5}<0,
\end{array}\end{cases}
\end{align}
where
\begin{equation}
\omega\pm\left(\zeta\right)=\frac{1}{6}(1-B)\pm\frac{\sqrt{2}}{3}[(1-B)(\frac{1}{2}+B)]^{1/2},\,B\equiv\sqrt{\frac{3}{2}\left(\zeta-\frac{1}{3}\right)}\;\in\left[0,1\right].
\end{equation}

The last two conditions for $\lambda_{2}$ must be satisfied for all
values of $\zeta\in[\frac{1}{3},1]$. Numerically, we consider 1000
steps in the interval of $\zeta$.

\vspace{0.5cm}

\textbf{The Higgs boson decays}

In this setup, the SM-like Higgs boson $h$ (the scalar with the mass $m_{h}=125.25\,\mathrm{GeV}$)
decays mainly into the fermions pairs $\tau^{+}\tau^{-},c\overline{c},b\overline{b}$
and the gauge bosons $WW^{*}$ and $ZZ^{*}$. The partial decay width
of the channel $h\to XX$ can be parametrized as $\Gamma(h\to XX)=\kappa_{X}^{2}\Gamma^{SM}(h\to XX)$,
where the coefficients,
\begin{equation}
\kappa_{\mathfrak{\mathrm{F}}}=\frac{g_{hff}^{GM}}{g_{hff}^{SM}}=\frac{c_{\alpha}}{c_{\beta}},\,\kappa_V=\frac{g_{hVV}^{GM}}{g_{hVV}^{SM}}=c_{\alpha}c_{\beta}-\sqrt{\frac{8}{3}}s_{\alpha}s_{\beta},\label{eq:K}
\end{equation}
represent the Higgs couplings modifiers with respect to the SM. This
allows us to write the total Higgs decay width as
\begin{equation}
\varGamma_{h}^{tot}=\Gamma_{h}^{SM}\sum_{X=SM}\kappa_{X}^{2}\mathcal{B}^{SM}(h\to XX),\label{eq:GM-Gamma}
\end{equation}
where $\varGamma_{h}^{SM}=4.08\,\mathrm{MeV}$~\cite{Workman:2022ynf}
and $\mathcal{B}^{SM}(h\to XX)$ are the SM values for total
decay width and the branching ratios for the Higgs boson, respectively.
Here, other decay channels like $h\to H_{3}H_{3}/H_{5}H_{5}$
could not be open due to the constraints on the charged scalar masses
$m_{H_{3}^{\pm}}^{2},\,m_{H_{5}^{\pm}}^{2}$ and $m_{H_{5}^{\pm\pm}}^{2}$.
The GM value for the Higgs boson (\ref{eq:GM-Gamma}) should lie in the
range~\cite{ATLAS},
\begin{equation}
2.1\,\mathrm{MeV}<\Gamma_{h}^{tot}<7.2\,\mathrm{MeV}.
\end{equation}

The signal strengths of the SM-like Higgs boson $h$ have been measured
in the LHC in various channels, where significant constraints are
established~\cite{Workman:2022ynf}. Here, one can translate these
constraints on the partial signal strength modifiers into bounds on
the GM Higgs couplings modifiers $\kappa_{X}$. In our analysis, we
consider only the gluon-gluon fusion $(ggF)$ Higgs production channel,
where the partial Higgs signal strength modifier of the channel $h\to XX$
can be simplified as
\begin{equation}
\mu_{XX}=\frac{\sigma(pp\to h)\times\mathcal{B}(h\to XX)}{\sigma^{SM}(pp\to h)\times\mathcal{B}^{SM}(h\to XX)}=\kappa_F^{2}\kappa_{X}^{2}\frac{\varGamma_{h}^{SM}}{\varGamma_{h}^{tot}},\label{eq:muXX}
\end{equation}
with $\sigma(gg\to h)\,[\sigma^{SM}(gg\to h)]$ is
the $ggF$ production cross section in the GM [SM] model. The constraints on the
invisible and undetermined channel are irrelevant here since they are closed due to the scalar masses $m_{3,5}>78\,\textrm{GeV}$, so ${\cal B}(h\to H_{3}^{\pm}H_{3}^{\mp},\,H_{5}^{\pm}H_{5}^{\mp})=0$. This means that the experimental measurements
of (\ref{eq:muXX}) will constraint significantly the coefficients
(\ref{eq:K}). Here, we consider the allowed values from all partial
Higgs strength modifiers within a $3\sigma$ range. The very recent
$1\sigma$ values are given in PDG by~\cite{Workman:2022ynf}
\begin{align}
\mu_{WW} & =1.19\pm0.12,~\mu_{ZZ}=1.01\pm0.07,~\mu_{b\overline{b}}=0.98\pm0.12,\nonumber \\
\mu_{\mu^{+}\mu^{-}} & =1.19\pm0.34,~\mu_{\tau^{+}\tau^{-}}=1.15_{-0.15}^{+0.16}.\label{muXX}
\end{align}

It is expected that (\ref{muXX}) put severe bounds on the Higgs coupling modifiers $\kappa_{F,V}$,
and consequently the mixing angles $\alpha$ and $\beta$.

\vspace{0.5cm}

\textbf{The electroweak precision tests}

The structure of the scalar-gauge interactions in the GM model makes
the constraints from the EWPTs very important. In the GM model, the
$T$ parameter estimation is problematic since it is divergent, but
the $S$ and $U$ parameters are calculable. Since the absolute value
of the $U$ parameter is found to be very small $<0.01$, we will
consider the constraint from the $S$ parameter by fixing the $U=0$.
The experimental values for the oblique parameter $S$ is extracted
for the SM Higgs mass $m_{h}=125.25\,\textrm{GeV}$, where we consider
the $2\sigma$ range in our numerical scan $S=0.05\pm0.11$~\cite{Baak:2014ora}.
The new contributions to the $S$ parameter~\cite{Hartling:2014aga}
in the GM model are given by
\begin{align}
\Delta S & =S_{GM}-S_{SM}=\frac{s_{W}^{2}c_{W}^{2}}{e^{2}\pi}\Big\{-\frac{e^{2}}{12s_{W}^{2}c_{W}^{2}}(\log\,m_{3}^{2}+5\log\,m_{5}^{2})+2\left|g_{ZhH_{3}^{0}}\right|^{2}f_{1}(m_{h},\,m_{3})\nonumber \\
 & +2\left|g_{Z\eta H_{3}^{0}}\right|^{2}f_{1}(m_{\eta},\,m_{3})+2(\left|g_{ZH_{5}^{0}H_{3}^{0}}\right|^{2}+2\left|g_{ZH_{5}^{+}H_{3}^{-}}\right|^{2})f_{1}(m_{5},\,m_{3})+\left|g_{ZZh}\right|^{2}\Big[\frac{f_{1}(m_{Z},\,m_{h})}{2m_{Z}^{2}}-f_{3}(m_{Z},\,m_{h})\Big]\nonumber \\
 & -\left|g_{ZZh}^{SM}\right|^{2}\Big[\frac{f_{1}(m_{Z},\,m_{h}^{SM})}{2m_{Z}^{2}}-f_{3}(m_{Z},\,m_{h}^{SM})\Big]+\left|g_{ZZ\eta}\right|^{2}\Big[\frac{f_{1}(m_{Z},\,m_{\eta})}{2m_{Z}^{2}}-f_{3}(m_{Z},\,m_{\eta})\Big]\nonumber \\
 & +\left|g_{ZZH_{5}^{0}}\right|^{2}\Big[\frac{f_{1}(m_{Z},\,m_{5})}{2m_{Z}^{2}}-f_{3}(m_{Z},\,m_{5})\Big]+2\left|g_{ZW^{+}H_{5}^{-}}\right|^{2}\Big[\frac{f_{1}(m_{W},\,m_{5})}{2m_{W}^{2}}-f_{3}(m_{W},\,m_{5})\Big]\Big\},\label{eq:dS}
\end{align}
with the functions $f_{1,3}$ and the couplings $g_{ZXY}$ are given
in Appendixes~\ref{app:func} and~\ref{app:coup},
respectively.

\vspace{0.5cm}

\textbf{The Higgs decays $h\to \gamma\gamma,\gamma Z$}

The Higgs decay into two photons or a photon and a Z gauge boson
are induced through a loop of charged particles. To estimate any new
physics effect on these Higgs decays, the ratios $R_{\gamma\gamma,\gamma Z}=\mathcal{B}(h\to \gamma\gamma,\gamma Z)/\mathcal{B}^{SM}(h\to \gamma\gamma,\gamma Z)$
are estimated and used to constrain the charged scalar masses and
their couplings to the Higgs boson. According to the latest data, we have
$R_{\gamma\gamma}=1.10\pm0.07$~\cite{Workman:2022ynf}. According
to the Feynman diagrams in Fig.~\ref{fig:Rgg}, the deviation of
$R_{\gamma\gamma}$ from unity, may come from many vertices such as
$\widetilde{g}\widetilde{g}h$, $t\bar{t}h$ and $W^{+}W^{-}h$ as
well due to new vertices involving new charged scalars.
\begin{figure}[h]
\includegraphics[width=0.99\textwidth]{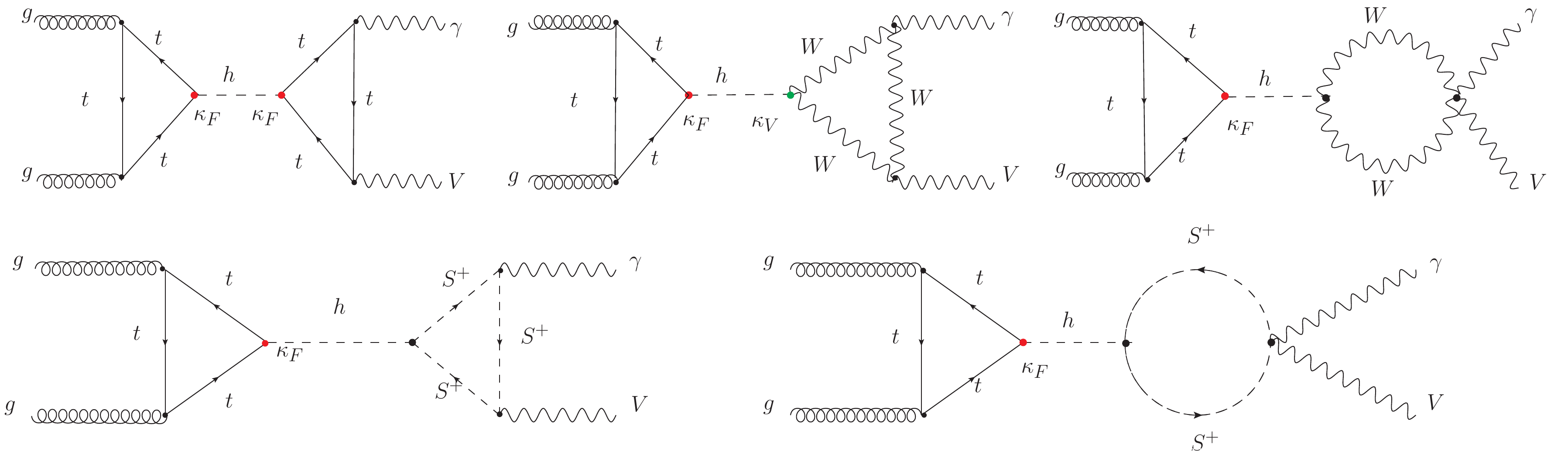} \caption{Feynman diagrams relevant to the Higgs decay $h\to \gamma~V$
($V=\gamma,Z$) at the LHC. The red and blue points refer to the vertices
that could be modified with respect to the SM by the factors $\kappa_F$
and $\kappa_V$, respectively.}
\label{fig:Rgg}
\end{figure}

From the diagrams in Fig.~\ref{fig:Rgg}, one finds the ratios
\begin{align}
R_{\gamma\gamma} & =\kappa_F^{2}\left|\frac{\frac{\upsilon}{2}\sum_{X}\frac{g_{hXX}}{m_{X}^{2}}Q_{X}^{2}A_{0}^{\gamma\gamma}(\tau_{X})+\kappa_VA_{1}^{\gamma\gamma}(\tau_{W})+\kappa_F\frac{4}{3}A_{1/2}^{\gamma\gamma}(\tau_{t})}{A_{1}^{\gamma\gamma}(\tau_{W})+\frac{4}{3}A_{1/2}^{\gamma\gamma}(\tau_{t})}\right|^{2},\label{eq:Rgg}\\
R_{\gamma Z} & =\kappa_F^{2}\left|\frac{\upsilon\sum_{X}\frac{g_{hXX}C_{ZXX}}{m_{X}^{2}}Q_{X}A_{0}^{\gamma Z}(\tau_{X},\lambda_{X})+\kappa_VA_{1}^{\gamma Z}(\tau_{W},\lambda_{W})+\kappa_F\frac{-6+16s_{\mathrm{w}}^{2}}{3s_{\mathrm{w}}c_{\mathrm{w}}}A_{1/2}^{\gamma Z}(\tau_{t},\lambda_{t})}{A_{1}^{\gamma Z}(\tau_{W},\lambda_{W})+\frac{-6+16s_{\mathrm{w}}^{2}}{3s_{\mathrm{w}}c_{\mathrm{w}}}A_{1/2}^{\gamma Z}(\tau_{t},\lambda_{t})}\right|^{2},\label{eq:RgZ}
\end{align}
where $X=H_{3}^{+},\,H_{5}^{+},\,H_{5}^{++}$ stands for all charged
scalars inside the loop diagrams, $Q_{X}$ is the electric charge
of the field $X$ in units of $|e|$, $\tau_{X}=4m_{X}^{2}/m_{h}^{2},~\lambda_{X}=4m_{X}^{2}/m_{Z}^{2}$;
and the functions $A_{i}^{\gamma\gamma,\gamma Z}$ and coefficients
$g_{hXX}$ and $C_{ZXX}$ are given in Appendixes~\ref{app:func}}
and\textit{~\ref{app:coup}}, respectively.

\vspace{0.5cm}

\textbf{Constraints from the production/decay of the heavy scalar
$\eta$}

After the discovery of the Higgs boson with $m_{h}=125.25\,\mathrm{GeV}$,
efforts have been devoted to search for heavy neutral scalar boson
through different channels over a wide mass range. Such results can
also be used to impose constraints on models with many neutral scalars
such as the GM model.

The two $CP-even$ eigenstates $h$ and $\eta$ are defined through a
mixing angle $\alpha$ and $(m_{h}<m_{\eta})$, where the light eigenstate
$h$ is identified to be the SM-like Higgs boson with the measured mass
$m_{h}=125.25\,\mathrm{GeV}$. Here, the heavy scalar $\eta$ has
similar couplings as the SM Higgs boson, but modified with the factors,
\begin{equation}
\zeta_V=\frac{g_{\eta VV}^{GM}}{g_{hVV}^{SM}}=s_{\alpha}c_{\beta}+\sqrt{\frac{8}{3}}c_{\alpha}s_{\beta},\,\zeta_F=\frac{g_{\eta FF}^{GM}}{g_{hFF}^{SM}}=\frac{s_{\alpha}}{c_{\beta}}.\label{zeta}
\end{equation}

The partial decay width of the heavy scalar $\eta$ into SM final
states can be written as $\Gamma(\eta\to X\overline{X})=\zeta_{X}^{2}\Gamma^{SM}(\eta\to X\overline{X})$,
where $\Gamma^{SM}(\eta\to X\overline{X})$ is the Higgs partial
decay width estimated at $m_{h}\to m_{\eta}$~\cite{Higgs}.
In addition, there exist other BSM decay channels like $\eta\to hh,H_{3}H_{3},H_{5}H_{5}$
when kinematically allowed. The partial decay width for these channels
is given by
\begin{equation}
\Gamma(\eta\to Y\overline{Y})=r_{Y}\frac{\left|g_{\eta Y\overline{Y}}\right|^{2}}{32\pi m_{\eta}}\sqrt{1-4\frac{m_{Y}^{2}}{m_{\eta}^{2}},}\label{eq:eta-width}
\end{equation}
with $Y=h,H_{3}^{0},H_{3}^{\pm},H_{5}^{0},H_{5}^{\pm},H_{5}^{\pm\pm}$,
$r_{h,H_{3}^{0},H_{5}^{0}}=1$ and $r_{H_{3}^{\pm},H_{5}^{\pm},H_{5}^{\pm\pm}}=2$.
Then, the heavy scalar $\eta$ total decay width can be written as
\begin{equation}
\varGamma_{\eta}^{tot}=\sum_{Y\ne SM}\Gamma(\eta\to Y\overline{Y})+\Gamma_{\eta}^{SM}\sum_{X=SM}\zeta_{Y}^{2}\mathcal{B}^{SM}(\eta\to X\overline{X}),\label{eq:eta-Gamma}
\end{equation}
where $\Gamma_{\eta}^{SM}$ and $\mathcal{B}^{SM}(\eta\to X\overline{X})$
are the Higgs total decay width and branching ratios estimated at
$m_{h}\to m_{\eta}$~\cite{Higgs}. Since the heavy scalar
$\eta$ decays into all SM final states, it can be searched at the
LHC via the processes: (1) $pp\to \eta\to \mathfrak{\ell\ell,\jmath\jmath,\mathrm{VV}}$
and $pp\to \eta\to hh$. For the first type, we include
the recent ATLAS analysis at $13\,\textrm{TeV}$ with $139\,\textrm{fb}^{-1}$
$pp\to \eta\to \tau\tau$~\cite{ATLAS:2020zms} and
$pp\to S\to ZZ$ via the channels $\ell\ell\ell\ell$
and $\ell\ell\nu\nu$~\cite{ATLAS:2020tlo}. In the other side, when
checking the bounds from the decay $pp\to \eta\to WW$,
one finds that the recent CMS analyses~\cite{CMS:2021klu} are not
convenient to use here, due to the considered large mass range $(m_{\eta}>1\,\textrm{TeV})$
in the analysis. For the second type, we use the recent ATLAS combination~\cite{ATLAS:2021nps}
that includes the analyses at $13\,\textrm{TeV}$ with $139\,\textrm{fb}^{-1}$
via the channels $hh\to b\bar{b}\tau\tau$~\cite{ATLAS:2021fet},
$hh\to b\bar{b}b\bar{b}$~\cite{ATLAS:2021ulo} and $hh\to b\bar{b}\gamma\gamma$
~\cite{ATLAS:2021jki}.

Here, we can take all the above mentioned analyses to constrain the
GM model parameters that are relevant to the heavy scalar $\eta$.
We define the cross section of the Heavy scalar $\eta$ in function
of the branching ratios and decay width as
\begin{equation}
\sigma(pp\to \eta)\times\mathcal{B}(\eta\to X\overline{X})=\zeta_F^{2}\zeta_{X}^{2}\frac{\Gamma_{SM}^{tot}(\eta)}{\Gamma^{tot}(\eta)}\sigma^{SM}(pp\to \eta)\times\mathcal{B}^{SM}(\eta\to X\overline{X}),\label{eq:XS}
\end{equation}
where $\mathcal{B}^{SM}(\eta\to X\overline{X})$ are the branching
ratios of the heavy scalar $\eta$ decaying into a pair of gauge bosons
or fermions via the ggF production mode of $\eta$, $\sigma(pp\to \eta)$
and $\sigma^{SM}(pp\to \eta)$ are the proton-proton collision
production cross section.

\vspace{0.5cm}

\textbf{LHC Constraints on the triplet and fiveplet Scalars}

Here, we implement some of the most stringent constraints, especially
the vector boson fusion (VBF) production of $H_{5}^{++}$ and the
Drell-Yan production of a neutral Higgs boson.

\vspace{0.5cm}
 \textbf{A. }\textbf{\textit{VBF $H_{5}^{++}\to W^{+}W^{+}\to $
like sign dileptons}}

The experimental bound on $s_{H}$ as a function of $m_{5}$ is constrained
by a CMS result of $35.9\,\textrm{fb}^{-1}$ of LHC run 2 (13 TeV)
data~\cite{CMS:2017fhs} for $m_{5}>200~\textrm{GeV}$, we assume that the
signal production cross section is proportional to $s_{H}^{2}$ where
\begin{equation}
(s_{H}^{limit})^{2}\times\mathcal{B}(H_{5}^{++}\to W^{+}W^{+})=(s_{H}^{CMS})^{2},\label{eq:VBF}
\end{equation}
with $(s_{H}^{CMS})^{2}$ is the bound presented at~\cite{CMS:2017fhs}
that corresponds to $\mathcal{B}(H_{5}^{++}\to W^{+}W^{+})=1$.

\vspace{0.5cm}

\textbf{B. }\textbf{\textit{Drell-Yan $H_{5}^{0}H_{5}^{\pm}$ with
$H_{5}^{0}\to \gamma\gamma$}}

Concerning the Drell-Yan production of $H_{5}^{0}H_{5}^{\pm}$ with
$H_{5}^{0}$, there exist two ATLAS searches for diphoton resonances in
the mass range $65<m_{5}<600~\textrm{GeV}$ using $20.3\,\textrm{fb}^{-1}$
of LHC run 1 (8 TeV) data~\cite{ATLAS:2014jdv} and of the $36.7\,\textrm{fb}^{-1}$ luminosity
of LHC run 2 (13 TeV) data in the mass range $200<m_{5}<2700~\textrm{GeV}$~\cite{ATLAS:2017ayi}.
The total cross sections at 8 TeV and 13 TeV for $H_{5}^{0}H_{5}^{+}$
and $H_{5}^{0}H_{5}^{-}$ are shown in~\cite{Ismail:2020zoz}. The
fiducial cross section is constrained by the following expression:
\begin{equation}
\sigma_{fiducial}=(\sigma_{H_{5}^{0}H_{5}^{+}}\times\epsilon_{+}+\sigma_{H_{5}^{0}H_{5}^{-}}\times\epsilon_{-})\times\mathcal{B}(H_{5}^{0}\to \gamma\gamma),\label{eq:Fiducial}
\end{equation}
where the efficiencies $\epsilon_{\pm}$ for $H_{5}^{0}H_{5}^{\pm}$
respectively, are shown in~\cite{Ismail:2020zoz}. As we will see later, only the 8 TeV constraints are relevant to (\ref{eq:Fiducial}) since the 13 TeV cross section values are 3 orders of magnitude suppressed with respect to the experimental bounds.

\vspace{0.5cm}

\textbf{The $b\to s$ transition bounds}

Since the charged triplet $H_{3}^{\pm}$ is partially coming from
the SM doublet as shown in (\ref{eq:Eigen}), then it couples to the
up and down quarks similar to the way the W gauge boson does. These
interactions lead to flavor violating processes such as the $b\to s$
transition ones, which depend only on the charged triplet mass $m_{3}$
and the mixing angle $\beta$. The current experimental value of the
$b\to s\gamma$ branching ratio, for a photon energy $E_{\gamma}>1.6~\textrm{GeV}$
 is $\mathcal{B}(\overline{B}\to X_{s}\gamma)_{exp}=(3.55\pm0.24\pm0.09)\times10^{-4}$,
while the two SM predictions are $\mathcal{B}(\overline{B}\to X_{s}\gamma)_{SM}=(3.15\pm0.23)\times10^{-4}$~\cite{Misiak:2006zs} and $\mathcal{B}(\overline{B}\to X_{s}\gamma)_{SM}=(2.98\pm0.26)\times10^{-4}$~\cite{Becher:2006pu}.
In our numerical scan, we consider the bounds on the $m_{3}$-$\upsilon_{\chi}$
plan shown in~\cite{Hartling:2014aga}.

\vspace{0.5cm}

\section{Numerical Analysis and Discussion\label{sec:NA}}

We perform a numerical scan over the parameter space of the GM model
and probe the effect of different theoretical and experimental constraints
on the parameter space. We require the light $CP-even$ scalar to be
the $125\,\textrm{GeV}$ SM-like Higgs boson and impose the constraints
from perturbativity, unitarity, boundness from below, the diphoton
Higgs decay, the Higgs total decay width, the Higgs signal strength
modifiers, the electroweak precision tests, the constraints from the
doubly charged Higgs bosons and Drell-Yan diphoton production, and
the indirect constraint from the $b\to s\gamma$ transition
processes.

We choose the model free parameters to be $\lambda_{2},\;\lambda_{4},\;m_{\eta},m_{3},\;m_{5},\,s_{\alpha}\;\textrm{and}\;s_{\beta}\equiv \sin\beta = 2\sqrt{2}\upsilon_{\xi}/\upsilon$,
which lie in the ranges,
\begin{equation}
78~\textrm{GeV}<m_{3}<1~\textrm{TeV},78~\textrm{GeV}<m_{5}<1.8~\textrm{TeV},\,m_{h}<m_{\eta}<1~\textrm{TeV},\,\left|\lambda_{2,4}\right|\leq10,\,\left|s_{\beta}\right|\leq1,\label{eq:range}
\end{equation}
where the triplet and fiveplet charged scalars are subject to a mass
lower bound from LEP~\cite{ALEPH:2013htx}. Here, the negative values of $s_{\beta}$ should be considered due to the following reason. In the GM model, we have $V(\Phi,\Delta,\mu_{1,2})=V(\Phi,-\Delta,-\mu_{1,2})$, and therefore all the mass matrix elements are also invariant under this transformation. However, since the scalar eigenstates are mixtures of the components of $\Phi$ and $\Delta$, the physical vertices that involves scalars are not invariant under $(\Phi,\Delta,\mu_{1,2})\to(\Phi,-\Delta,-\mu_{1,2})$. This means that any two BPs with the same input parameters but with opposite signs of $(\pm s_{\beta},\pm\mu_{1,2})$ are physically different. This makes the negative $s_{\beta}$ values in (\ref{eq:range}) independent parameter space that should not be ignored.

\begin{figure}[h]
\includegraphics[width=8.3cm,height=5cm]{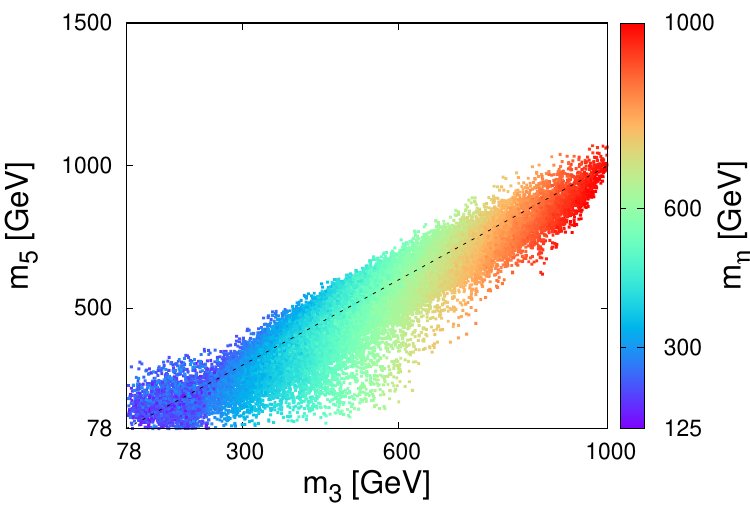}~
\includegraphics[width=8.3cm,height=5cm]{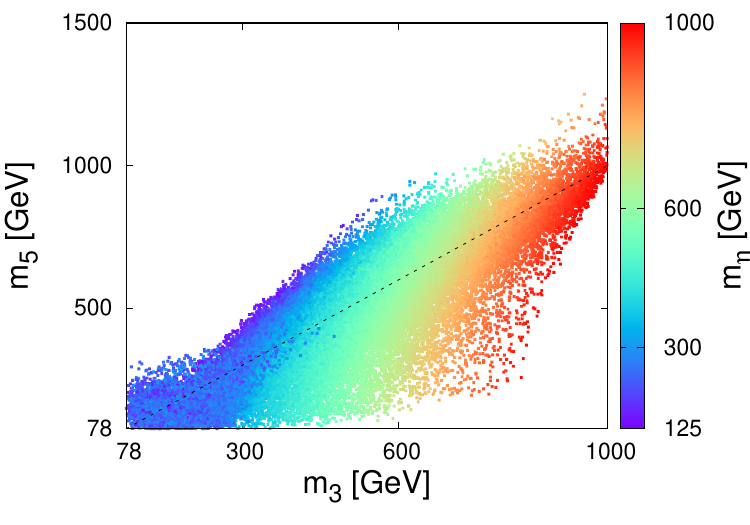}
\caption{The masses for triplet, fiveplet and singlet $\eta$ estimated in
the GM model by considering the basic theoretical and experimental
constraints with (left) and without (right) the condition of the EW
vacuum to be the deepest (\ref{eq:Vac}).}
\label{fig:Vac}
\end{figure}

In order to check whether there exist wrong vacua that are deeper
than the EW one $(\upsilon_{\phi},\,\sqrt{2}\upsilon_{\xi},\upsilon_{\xi})$,
we show in Fig.~\ref{fig:Vac} the scalar mass ranges with (left)
and without (right) the condition (\ref{eq:Vac}).

From the 58.5k BPs, 35k BPs fulfill the condition
(\ref{eq:Vac}). This means that almost 40 \% of the parameter space considered
in the literature are excluded by the fact that the EW vacuum $(\upsilon_{\phi},\,\sqrt{2}\upsilon_{\xi},\upsilon_{\xi})$
is not the deepest one. Clearly, when considering all the theoretical and experimental constraints except the condition (\ref{eq:Vac}), the fiveplet and the singlet $\eta$ masses can
reach the values $m_{5}=1.25\,\textrm{TeV}$ and $m_{\eta}=1\,\textrm{TeV}$, respectively for the triplet maximal mass value $m_{3}=1\,\textrm{TeV}$. However, when considering
the constraint (\ref{eq:Vac}), the fiveplet mass ranges
get shrunk as $m_{5}<1.1~\textrm{TeV}$. This requires a full reanalysis of different
phenomenological aspects of this model. The viable parameter space
in Fig.~\ref{fig:Vac}-right is a consequence of a combination of
the theoretical and experimental constraints mentioned above.

\begin{figure}[h]
\includegraphics[width=8.3cm,height=5cm]{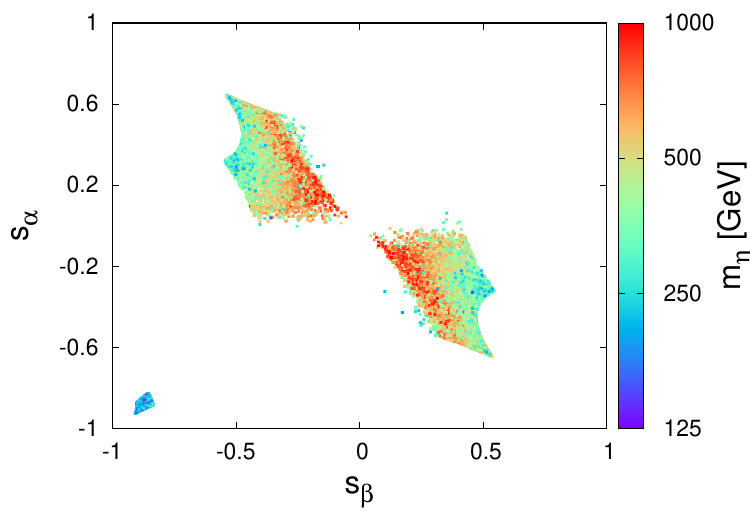}~
\includegraphics[width=8.3cm,height=5cm]{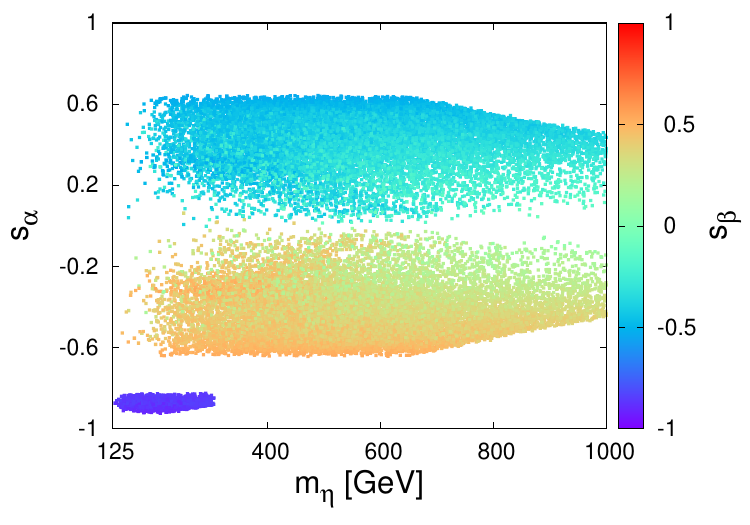}\\
\includegraphics[width=8.3cm,height=5cm]{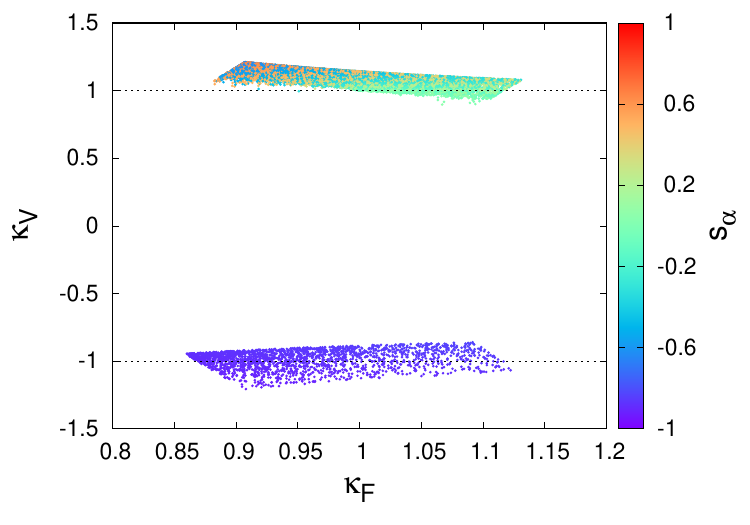}~
\includegraphics[width=8.3cm,height=5cm]{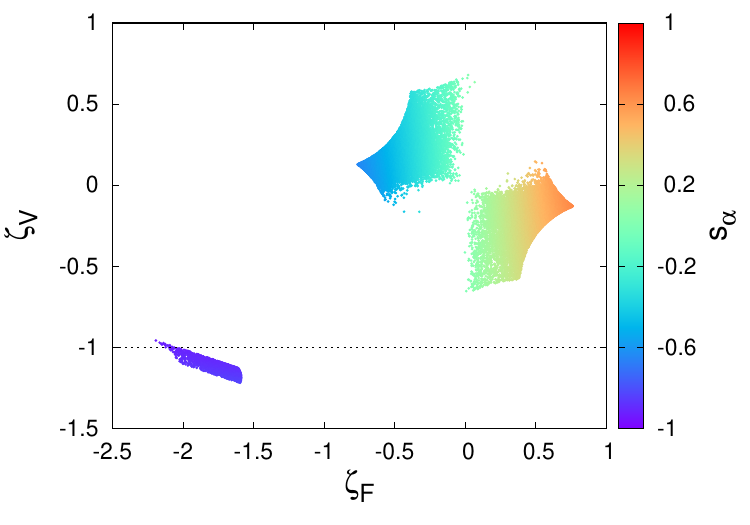}
\caption{Different physical observables estimated in the GM model by considering
the theoretical and experimental constraints, i.e., the BPs used to
produce Fig.~\ref{fig:Vac}-left.}
\label{fig:Pr}
\end{figure}

In what follows, we will consider only the 35k viable
BPs in our analysis, as shown in Fig.~\ref{fig:Pr}

From Fig.~\ref{fig:Pr}, one has to mention that the parameter space is well constrained and split
into three isolated islands in the plans of $\{s_{\beta},s_{\alpha}\}$, $\{s_{\alpha},m_{\eta}\}$ and $\{\zeta_F,\zeta_V\}$; and into two islands in the plans of $\{\kappa_F,\kappa_V\}$. For instance,
the three islands correspond to the ranges \{-0.92$<s_{\beta}<$-0.83,~-0.92$<s_{\alpha}<$-0.81\},
\{-0.54$<s_{\beta}<$-0.05 ,~0.01$<s_{\alpha}<$0.64\} and \{0.04$<s_{\beta}<$0.54,~-0.64$<s_{\alpha}<$0.03\},
respectively. According to the bottom-right panel in Fig.~\ref{fig:Pr}, the
$\kappa$'s values for the two islands are \{-1.21$<\kappa_V<$-0.85,~0.86$<\kappa_F<$1.12\} and \{0.9$<\kappa_V<$1.23,~0.88$<\kappa_F<$1.13\},
respectively. While, the corresponding $\zeta$'s ranges are $\{-1.22<\zeta_V<-0.97,~-2.15<\zeta_F<-1.59\}$, $\{-0.09<\zeta_V<0.66,~-0.75<\zeta_F<-0.02\}$ and $\{-0.65<\zeta_V<0.14,~0.04<\zeta_F<0.75\}$,
for the three islands, respectively. Here, the shape of all islands is dictated by the combination of all the above mentioned constraints, however, some of the constraints could have the dominant impact on such a region. For instance, the shape of the isolated islands is mainly dictated by the bounds from $b\to s$.

The Higgs coupling modifier $\kappa_V$ is very constrained and
could have both signs, while the $\kappa_F$ deviation with respect
to the SM can reach 13 \%. These deviations of $\kappa_{F,V}$ form
the SM are possible due to the strength of the bounds from some experimental
constraints, such as the diphoton Higgs decay, the bounds on the
total Higgs decay width and the Higgs signal strength modifiers. Unlike
most of the SM extensions that involve a heavy scalar whose couplings
to the fermions and gauge bosons are similar to those of the SM-like
Higgs bosons, the scaling factor could have values larger than unity $|\zeta_F|>1$. The reason of the significant deviation
of the factors $\zeta_V,\kappa_V$ from unity, could be the factor
$\sqrt{8/3}$, in addition to the sine and cosine in the denominator
in (\ref{zeta}) and (\ref{eq:K}). These values are very similar
to the results obtained in~\cite{Ismail:2020zoz} for the region
of positive $\kappa_V$ due to the stringent constraints from
the $b\to s$ transition bounds. However, we got another region
with negative $\kappa_V$ values that is not mentioned in~\cite{Ismail:2020zoz},
as it is allowed all the constraints considered in our scan of the
full free parameters ranges (\ref{eq:range}).

In the majority of SM scalar extensions where the heavy scalar $\eta$ couplings to the fermions and
gauge bosons are much smaller than the SM values ($|\zeta_{F,V}|\ll1$). This makes these models in agreement with all the negative searches of a heavy resonance. But in the GM model, the situation is different, i.e., $\zeta_{F,V}$ are not suppressed, and these negative
searches could play a key role to exclude most of the parameter space
as will be seen next.

\begin{figure}[h]
\includegraphics[width=8.3cm,height=5cm]{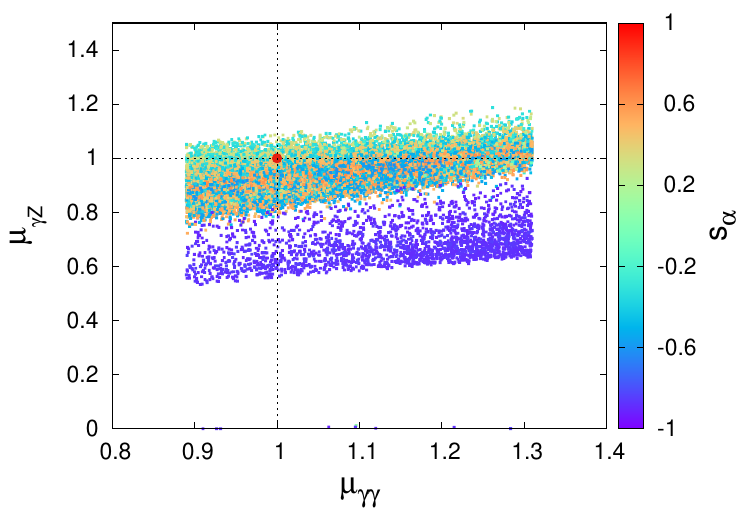}~
\includegraphics[width=8.3cm,height=5cm]{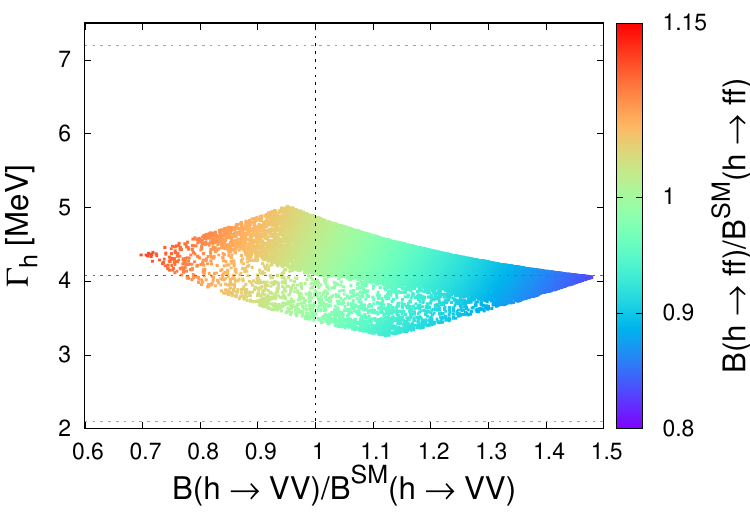}
\caption{Left: the ratio $R_{\gamma Z}$ in function of $R_{\gamma\gamma}$,
where the palette shows the sine of the mixing angle $\alpha$. Right:
the SM-like Higgs total decay width versus Higgs branching ratio to
gauge bosons scaled by its SM value. The palette shows the Higgs branching
ratio to fermions scaled by its SM value; and the dashed line at $\Gamma_h=4.08\,\textrm{MeV}$ corresponds to the SM value, while the experimentally allowed values are shown by the dashed lines at $2.1\,\textrm{MeV}$ and $7.2\,\textrm{MeV}$~\cite{ATLAS}.}
\label{fig:higgs}
\end{figure}

In Fig.~\ref{fig:higgs}, we show the ratios $R_{\gamma\gamma}$
and $R_{\gamma Z}$ for the SM-like Higgs boson (left) and the Higgs total
decay width versus its branching ratios (right).

From Fig.~\ref{fig:higgs}-left, while the values of $R_{\gamma\gamma}$
are constrained by the current LHC data~\cite{Workman:2022ynf},
the ratio $R_{\gamma Z}$ is modified drastically with respect to
the SM, it could be reduced by $\sim -45\%$ as it could be $\sim 18\%$ enhanced with respect to the SM. There are few BPs where $R_{\gamma Z}$ is almost null, which correspond to some specific values of $\kappa_{F,V}$, where a possible cancellation could occur between different terms in (\ref{eq:RgZ}). From the right panel, one learns that the Higgs
decays into gauge bosons and fermions can be reduced/enhanced by $-70-150\%$
and $-90-110\%$, respectively. Therefore, more precise Higgs
measurements will tighten these ranges and constraint more the parameter
space. For the considered parameter space, the oblique parameter given
in (\ref{eq:dS}) takes the values $-0.17<\Delta S<0.25$.

\begin{figure}[h]
\includegraphics[width=0.33\textwidth]{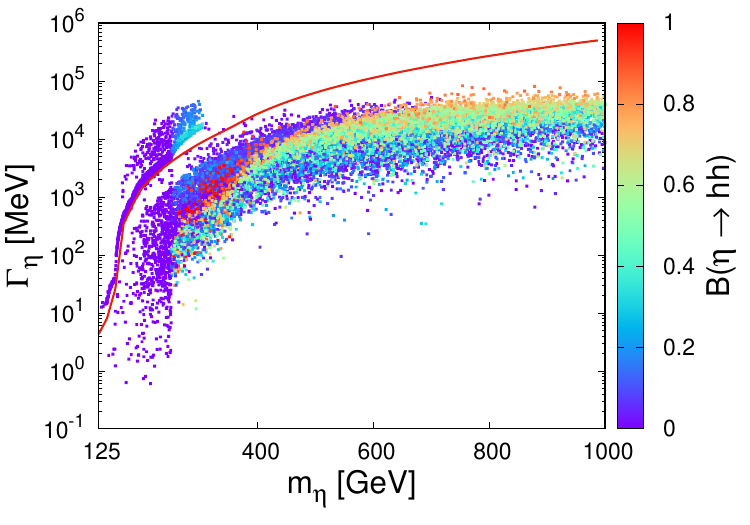}~\includegraphics[width=0.33\textwidth]{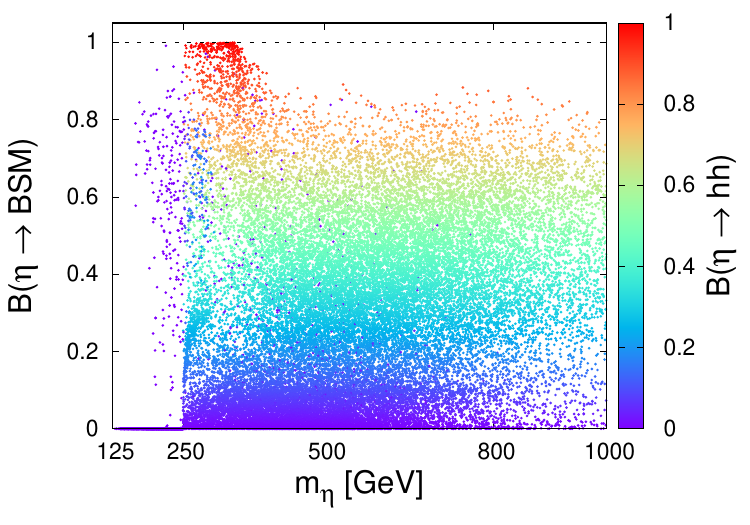}~\includegraphics[width=0.33\textwidth]{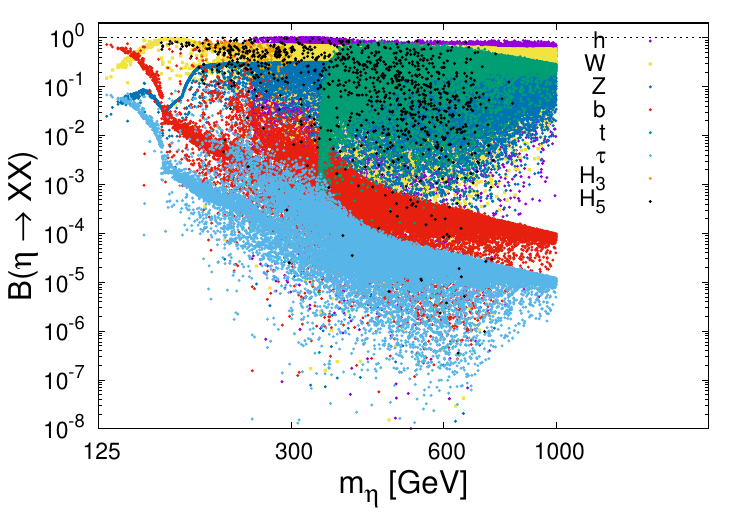}
\caption{Left: the total decay width of the scalar $\eta$ in function
of its mass $m_{\eta}$, where the palette shows its di-Higgs branching
ratio. The red curve represents the total decay $\Gamma_{\eta}$ estimated
in the SM~\cite{Higgs}, i.e., with $s_{\alpha}=1$ and $B_{BSM}=0$.
Middle: the BSM branching ratio $BSM=h,H_{3},H_{5}$ versus $m_{\eta}$,
where the palette shows the di-Higgs branching ratio. Right: the
branching ratios ${\cal B}(\eta\to XX)$ versus $m_{\eta}$.}
\label{fig:eta1}
\end{figure}

In Fig.~\ref{fig:eta1}, we present some observables relevant to the
heavy scalar $\eta$ versus its mass. In the left panel we show its
total decay width and its invisible and undetermined branching
fractions in the middle panel, while the SM branching ratios are shown in the right panel.

One has to mention that the singlet scalar $\eta$ total decay width
could be either 2 orders of magnitude smaller or larger than SM
estimated value as shown in Fig.~\ref{fig:eta1}-left. This can
be understood due the possible significant deviation of the factors
$\zeta_{F,V}$ from unity, in addition to possible large values for
the possible partial decay widths for $\eta\to hh,H_{3}H_{3},H_{5}H_{5}$.
According to Fig.~\ref{fig:eta1}-middle, one notices that the
BSM channels could be dominant for $m_{\eta}>160~\textrm{GeV}$. Here,
one notes that the BSM branching ratios are dominant by $\eta\to H_{3}H_{3}$
and $\eta\to H_{5}H_{5}$ in the region of mass $145\,\textrm{GeV}<m_{\eta}<250\,\textrm{GeV}$
but when $m_{\eta}>250\,\textrm{GeV}$ the BSM branching ratio is
dominant by $\eta\to hh$.
Clearly from Fig.~\ref{fig:eta1}-right, one remarks that the branching
ratios $\mathcal{B}(\eta\to WW,\,ZZ,\,b\overline{b},\,\tau\tau,\,tt)$
are comparable to their SM corresponding values~\cite{Higgs} for
a large portion of the BPs.

\begin{figure}[h]
\includegraphics[width=0.48\textwidth]{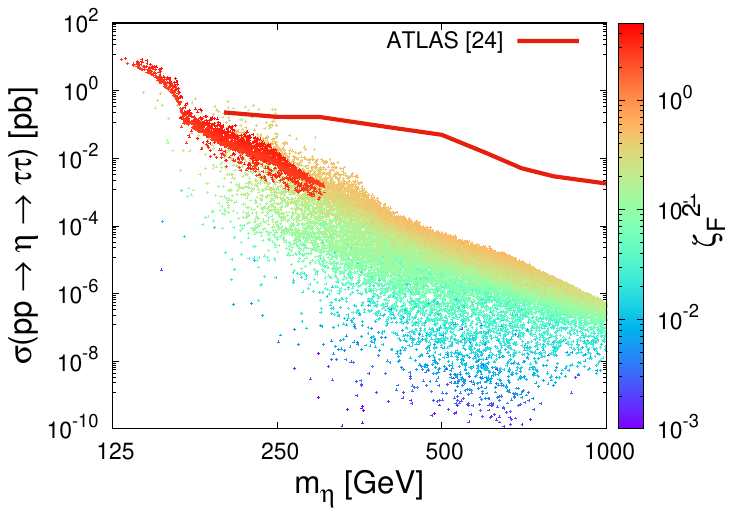}~
\includegraphics[width=0.48\textwidth]{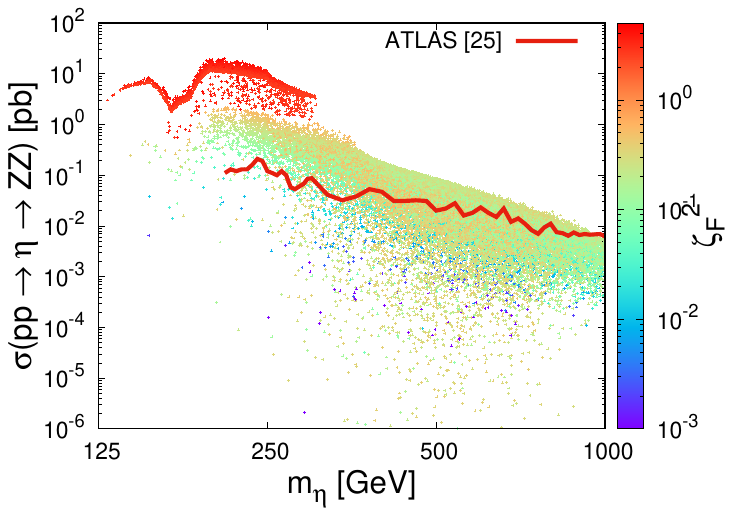}
\caption{The resonant production cross section $pp\to \eta\to \tau\tau$
(left) and $pp\to \eta\to ZZ$ (right) as a function
of the heavy scalar mass $m_{\eta}$, where the palette shows the
square of the scaling factor $\zeta_F$. The red curves represent
the corresponding experimental bounds from ATLAS~\cite{ATLAS:2020zms,ATLAS:2020tlo}.}
\label{fig:eta2}
\end{figure}

In Fig.~\ref{fig:eta2}, we show the resonant production cross section of the
heavy scalar $\eta$ compared to the experimental bounds in the channels
$\tau\tau$ (left) and $ZZ$ (right).

From Fig.~\ref{fig:eta2}, the
experimental bounds from the negative searches for a heavy resonance
in the channels $\tau\tau$ and $ZZ$ exclude significant part of
the parameter space. However, more regions in the parameter space
will be excluded if the future searches for a heavy resonance would
consider the mass range $125-200~\textrm{GeV}$. For the $\eta \to ZZ$ constraint, if one extrapolates the bound into small $m_{\eta}$ values, one learns that all the BPs with $\zeta_F^2>0.6$ are excluded.

\begin{figure}[h]
\includegraphics[width=8.3cm,height=5cm]{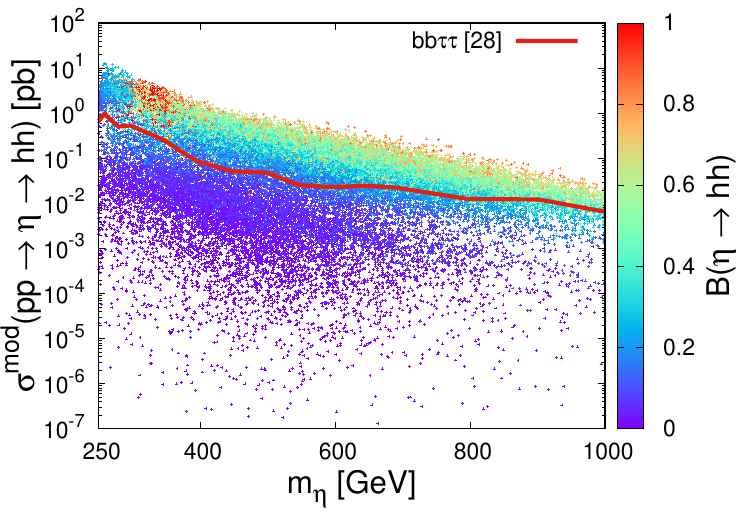}~
\includegraphics[width=8.3cm,height=5cm]{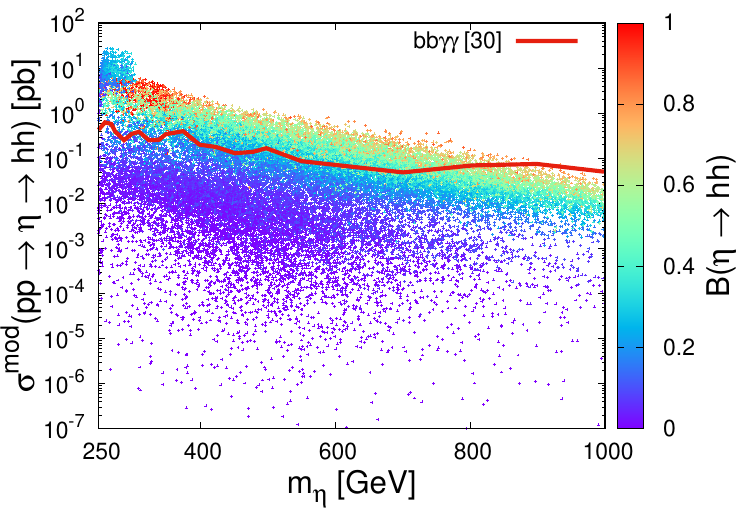}
\caption{The $hh$ production cross section (\ref{mod}) as a function of $m_{\eta}$
from the combination of $hh\to b\bar{b}\tau\tau$~\cite{ATLAS:2021fet},
(left) and via $hh\to b\bar{b}\gamma\gamma$~\cite{ATLAS:2021jki}
(right).}
\label{fig:XS}
\end{figure}

Concerning the resonant production $\eta\to hh$, the production
cross section can not be directly compared to the experimental bounds
in the channels $hh\to b\bar{b}\tau\tau$~\cite{ATLAS:2021fet},
$hh\to b\bar{b}b\bar{b}$~\cite{ATLAS:2021ulo} and $hh\to b\bar{b}\gamma\gamma$~\cite{ATLAS:2021jki},
since these analyses have been performed by taking into account the
SM Higgs branching ratio. Therefore, the modified cross section
\begin{equation}
\sigma^{mod}(pp\to \eta\to hh)=\sigma^{GM}(pp\to \eta\to hh)\times\frac{\mathcal{B}(h\to X_{1}\bar{X}_{1})\mathcal{B}(h\to X_{2}\bar{X}_{2})}{\mathcal{B}^{SM}(h\to X_{1}\bar{X}_{1})\mathcal{B}^{SM}(h\to X_{2}\bar{X}_{2})},\label{mod}
\end{equation}
is the relevant quantity to be compared with the experimental bounds~\cite{ATLAS:2021fet,ATLAS:2021ulo,ATLAS:2021jki}
in the channel $hh\to X_{1}\bar{X}_{1}X_{2}\bar{X}_{2}$. In Fig.~\ref{fig:XS}, we show the modified cross section (\ref{mod})
as a function of the heavy scalar mass from the combination of $hh\to b\bar{b}\tau\tau$
and $hh\to b\bar{b}\gamma\gamma$ for the BPs with $m_{\eta}>250~\textrm{GeV}$,
where the palette shows the branching ratio of $\eta\to hh$.

From Fig.~\ref{fig:XS}, one learns that the majority of the BPs
with $m_{\eta}>250~\textrm{GeV}$ are excluded by the experimental
bounds~\cite{ATLAS:2020zms,ATLAS:2020tlo,CMS:2021klu,ATLAS:2021nps,ATLAS:2021fet,ATLAS:2021ulo,ATLAS:2021jki}.
One has to mention that the di-Higgs negative searches are used to
set some limits on the triple Higgs couplings and to constrain the
scalar sector in many multiscalar SM extensions, but here in the
GM model, the resonant $\eta\to hh$ experimental bounds are
very efficient in excluding large part of the parameter space. This
point will be investigated in details in a future work~\cite{next}.

\begin{figure}[h]
\includegraphics[width=8.3cm,height=5cm]{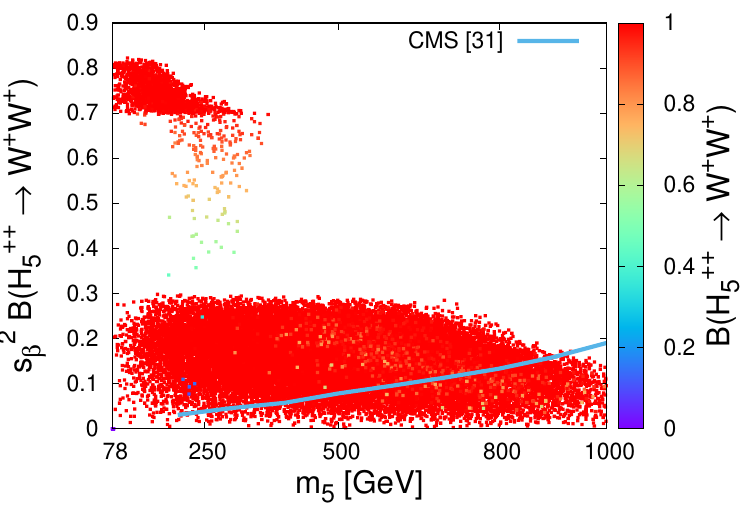}~
\includegraphics[width=8.3cm,height=5cm]{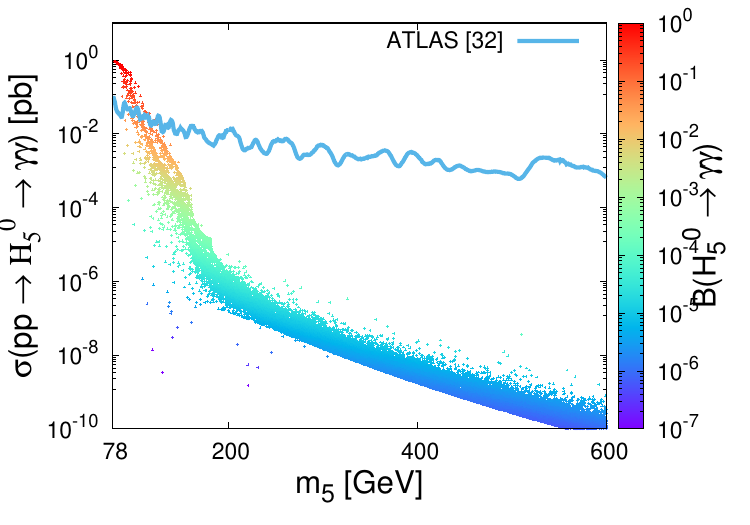}
\caption{Left: the quantity $s_{\beta}^{2}\times\mathcal{B}(H_{5}^{++}\to W^{+}W^{+})$
in function of $m_{5}$, where the palette shows the branching ratio
$\mathcal{B}(H_{5}^{++}\to W^{+}W^{+})$. The blue curve represents
the experimental bounds from CMS~\cite{CMS:2017fhs}. Right: the
cross section of the diphoton production at 8 TeV, where the palette
shows the corresponding branching ratio. The blue curve shows the experimental
bound~\cite{ATLAS:2014jdv}. Here, the BPs with $m_5>600~\textrm{GeV}$ are not considered since by the experimental bound~\cite{ATLAS:2014jdv} were established only for $m_5<600~\textrm{GeV}$.}
\label{fig:VBF}
\end{figure}

Here, in Fig.~\ref{fig:VBF} we show the effect of the constraints
from the doubly charged Higgs bosons and Drell-Yan diphoton production on different
observables like $s_{\beta}^{2}\times\mathcal{B}(H_{5}^{++}\to W^{+}W^{+})$
and the cross section of the diphoton production at 8 TeV which are
plotted in function of $m_{5}$ and the corresponding branching ratio
in the palette. One has to mention that it is worthless to show
the cross section $pp\to H_{5}^{0}\to\gamma\gamma$ at 13 TeV since
the existing experimental bounds are given for the $m_{5}$ range~\cite{ATLAS:2017ayi},
that it is already excluded by previous constraints.

One notices from Fig.~\ref{fig:VBF}-left that the branching ratio $\mathcal{B}(H_{5}^{++}\to W^{+}W^{+})$ value does not play an important role in excluding the BPs by the experimental bounds~\cite{CMS:2017fhs}; however, the mixing value $s_{\beta}$ does. From Fig.~\ref{fig:VBF}-right, one remarks that most of
the diphoton scalar negative searches exclude most of the BPs with
$\mathcal{B}(H_{5}^{0}\to \gamma\gamma)>0.09$, which is in
good agreement with the experimental bound~\cite{ATLAS:2014jdv}.

\begin{figure}[h]
\includegraphics[width=8.3cm,height=5cm]{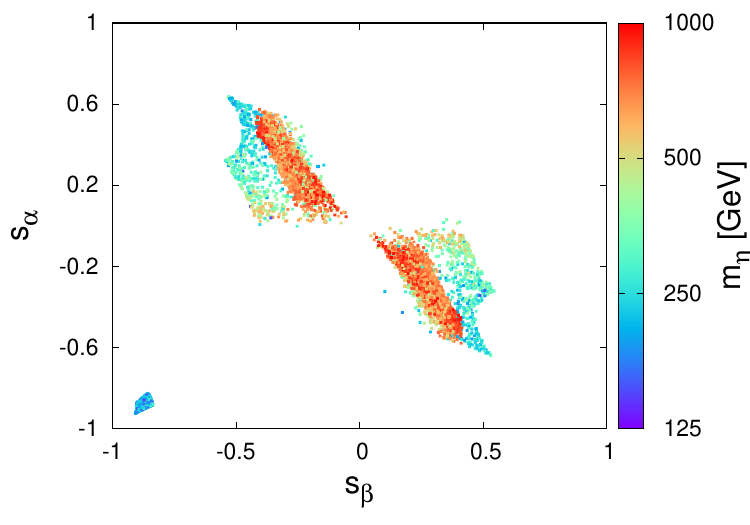}~
\includegraphics[width=8.3cm,height=5cm]{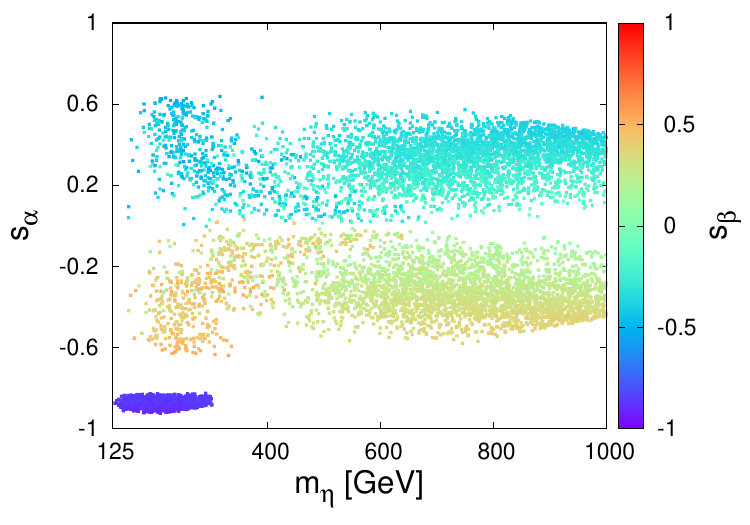}\\
\includegraphics[width=8.3cm,height=5cm]{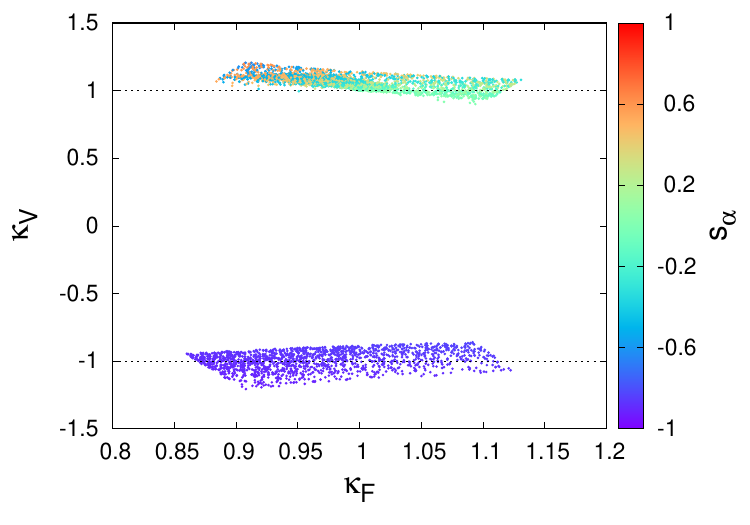}~
\includegraphics[width=8.3cm,height=5cm]{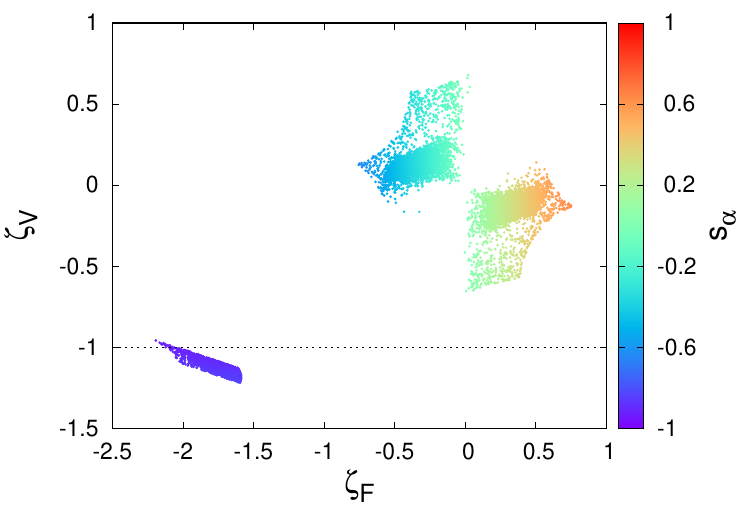}
\caption{The physical observables that are presented in Fig.~\ref{fig:Pr}
reproduced using only BPs that are in agreement with the recent ATLAS
and CMS experimental bounds~\cite{ATLAS:2020zms,ATLAS:2020tlo,CMS:2021klu,ATLAS:2021nps,ATLAS:2021fet,ATLAS:2021ulo,ATLAS:2021jki};
in addition to the constraints from the doubly charged Higgs bosons
and Drell-Yan di-photon production~\cite{CMS:2017fhs,ATLAS:2014jdv,ATLAS:2017ayi}
as well as the indirect constraints from $b\to s\gamma$ and
the LHC measurements on the Higgs strengths modifiers.}
\label{fig:Ch}
\end{figure}

In Fig.~\ref{fig:Ch}, we reproduce the physical observables shown
in Fig.~\ref{fig:Pr} by considering only the BPs that are in agreement
with all the above mentioned experimental bounds~\cite{ATLAS:2020zms,ATLAS:2020tlo,CMS:2021klu,ATLAS:2021nps,ATLAS:2021fet,ATLAS:2021ulo,ATLAS:2021jki}
fulfill the constraints from doubly charged Higgs boson and Drell-Yan
diphoton production, the indirect constraints from the $b\to s$
transition processes and the LHC measurements on the Higgs strengths
modifiers.

From the 35k BPs considered in our analysis, 74.5 \% are excluded by
the above combined constraints, where the majority of BPs correspond
to $m_{\eta}<2m_{h}$. However, most of them are not excluded due
the absence of the experimental bounds for $m_{5}<200~\textrm{GeV}$. By comparing Fig.~\ref{fig:Ch} with Fig.~\ref{fig:Pr}, one has to mention that these constraint do not change the shape of the islands described previously.

Before concluding this debate, it is essential to discuss the impact of future measurements
at the HL-LHC on this model. The primary objectives of the HL-LHC include enhancing measurements related to the
125 GeV Higgs boson's couplings, decays, and the search for heavy Higgs particles. Additionally, it offers an
important opportunity to test some BSM theories. In a study by Li~\cite{Li:2018jns}, the possibility of observing the type-II seesaw doubly charged scalar was investigated. They obtained a mass upper bound of $655~\textrm{GeV}$, which is irrelevant to the doubly charged scalar in the current model. According to the projections for Higgs property measurements~\cite{TheATLAScollaboration:2014ewu}, it is expected that the various Higgs scaling factors and, consequently, the signal strength modifiers in (\ref{eq:muXX}) will be measured with significantly improved precision. This will result in narrower experimentally allowed ranges for the scaling factors $\kappa_{F,V}$ as shown in (\ref{muXX}), leading to the exclusion of a significant portion of the parameter space.

\section{Conclusion\label{sec:Conclusion}}

In this work, we have studied the scalar potential of the GM model
that preserves custodial SU(2) symmetry. We have considered the theoretical
and experimental constraints on the parameter space such as the tree-level
unitarity, the potential boundness from below, avoiding possibly deeper
wrong minima, the electroweak precision tests, the Higgs total decay
width and diphoton decay, and the Higgs strength modifiers, the negative searches on the doubly charged Higgs bosons and the Drell-Yan diphoton production, as well as the indirect constraints from the $b\to s$ transition processes; in addition to the direct searches for additional heavy Higgs resonances.

We performed a numerical scan based on all the above-mentioned theoretical and experimental constraints, and we found that the possible unwanted minima that could be deeper than the EW vacuum excludes about 40 \% of the parameter space that fulfills the above mentioned constraints. On top of that, we noticed that the above constraints dictate a clear shape on the model parameter of three separated islands in the plans of $\{s_{\beta},s_{\alpha}\}$, $\{s_{\alpha},m_{\eta}\}$ and $\{\zeta_F,\zeta_V\}$, and two islands in the plans of $\{\kappa_F,\kappa_V\}$. The couplings of the Higgs boson to the gauge bosons and fermions lie in the ranges \{-1.21$<\kappa_V<$-0.85,~0.86$<\kappa_F<$1.12\} and \{0.9$<\kappa_V<$1.23,~0.88$<\kappa_F<$1.13\}, respectively. However, the scaling factors of the heavy scalar $\eta$ in the GM $\zeta_F,_V$ lie in the ranges \{-1.22$<\zeta_V<$-0.97,~-2.15$<\zeta_F<$-1.59\},
\{-0.09$<\zeta_V<$0.66,~-0.75$<\zeta_F<$-0.02\} and \{-0.65$<\zeta_V<$0.14,~0.04$<\zeta_F<$0.75\}, respectively. Here, an isolated islands in the plans of that was supposed to exist was excluded by the $b\to s$ bound. The shape of the isolated islands as shown in the plans of $\{s_{\beta},s_{\alpha}\}$, $\{s_{\alpha},m_{\eta}\}$, $\{\zeta_F,\zeta_V\}$ and $\{\kappa_F,\kappa_V\}$ is dictated by the combination of the bounds of the Higgs signal strength modifiers and the Higgs total decay width; in addition to the Higgs diphoton decay.

We have also imposed the constraints
from the negative searches of both doubly charged Higgs bosons in
the VBF channel and Drell-Yan diphoton production, where we found
that a significant part of the parameter space is excluded by the
CMS bound on $s_{\beta}^{2}\times\mathcal{B}(H_{5}^{++}\to W^{+}W^{+})$~\cite{CMS:2017fhs}.
Here, it has been found that the branching ratio of $H_{5}^{++}\to W^{+}W^{+}$ does not play an important role in allowing/excluding any BP, but
the mixing $s_{\beta}$ does. Unfortunately, the recent bounds from CMS~\cite{CMS:2017fhs}
and ATLAS~\cite{ATLAS:2017ayi} do not cover the mass range $m_{5}<200~\textrm{GeV}$,
which makes a large part of the parameter space unconstrained by this
severe bound. It will be interesting if future analyses would consider this mass range.

The indirect constraints from the $b\to s$ transition processes
are also applied and put constraints on the two parameters $m_{3}$
and $\upsilon_{\xi}$ only. We found also that the recent LHC measurements
on the Higgs strengths modifiers impose strong constraints on the parameter
space, especially the Higgs coupling modifiers $\kappa_{F,V}$. In
fact, the direct searches generally provide more strict constraints
on the GM model parameter space and open the possibility of a discovery
as these searches would be improved within the current/future LHC
data. We have imposed also the recent ATLAS and CMS negative searches
for the heavy scalar $\eta$ in different channels. We found that
the channel $\eta\to hh$ is very useful to exclude most of
the parameter space, while, other channels are less efficient since
the mass range $125~\textrm{GeV}<m_{\eta}<200~\textrm{GeV}$ is not
covered by most of the searches. Clearly, future searches and more
precise measurements will tighten the parameter space of the GM model.

\appendix

\section{FUNCTIONS\label{app:func}}

The loop functions used in (\ref{eq:dS}) are given by

\begin{align}
f_{1}(x,\,y) & =\frac{5(y^{6}-x^{6})+27(x^{4}y^{2}-x^{2}y^{4})+12(x^{6}-3x^{4}y^{2})\log x+12(3x^{2}y^{4}-y^{6})\log y}{36(y^{2}-x^{2})^{3}},\nonumber \\
f_{3}(x,\,y) & =\frac{x^{4}-y^{4}+2x^{2}y^{2}(\log y^{2}-\log x^{2})}{2(x^{2}-y^{2})^{3}},
\end{align}
while those used in (\ref{eq:Rgg}) and (\ref{eq:RgZ}) are given
by~\cite{Djouadi:2005gj}
\begin{align}
A_{1}^{\gamma\gamma}(\tau) & =2+3\tau+3\tau(2-\tau)f(\tau),~A_{1/2}^{\gamma\gamma}(\tau)=-2\tau[1+(1-\tau)f(\tau)],~A_{0}^{\gamma\gamma}(\tau)=\tau[1-\tau f(\tau)],\nonumber \\
A_{1}^{\gamma Z}(\tau,\lambda) & =-\cot\theta_{W}\big(4(3-\tan^{2}\theta_{W})I_{2}(\tau,\lambda)+[(1+\frac{2}{\tau})\tan^{2}\theta_{W}-(5+\frac{2}{\tau})]I_{1}(\tau,\lambda))\big),\nonumber \\
A_{1/2}^{\gamma Z}(\tau,\lambda) & =I_{1}(\tau,\lambda)-I_{2}(\tau,\lambda),~A_{0}^{\gamma Z}(\tau,\lambda)=I_{1}(\tau,\lambda)\nonumber \\
I_{1}(a,\,b) & =\frac{ab}{2(a-b)}+\frac{a^{2}b^{2}}{2(a-b)^{2}}[f(a)-f(b)]+\frac{a^{2}b}{(a-b)^{2}}[g(a)-g(b)],\nonumber \\
I_{2}(a,\,b) & =-\frac{ab}{2(a-b)}[f(a)-f(b)],
\end{align}
with
\begin{equation}
f(\tau)=\begin{cases}
[\arcsin\big(\sqrt{\frac{1}{\tau}}\big)]^{2} & if\;\tau\geq1,\\
-\frac{1}{4}\Big[\log\big(\frac{1+\sqrt{1-\tau}}{1-\sqrt{1-\tau}}\big)-i\pi\Big]^{2} & \,\,\,\,if\;\tau<1
\end{cases},\,\,\,\,\,\,\,\,\,\,\,g(\tau)=\begin{cases}
\sqrt{\tau-1}[\sin^{-1}(\sqrt{\frac{1}{\tau}})] & if\;\tau\geq1,\\
\frac{1}{2}\sqrt{\tau-1}[\log(\frac{\eta_{+}}{\eta_{-}})-i\pi] & if\;\tau<1.
\end{cases}
\end{equation}

\section{COUPLINGS\label{app:coup}}

Here, we give the couplings used in different observables definitions.
The couplings that are used in (\ref{eq:dS}) are
\begin{align}
g_{ZhH_{3}^{0}} & =-i\sqrt{\frac{2}{3}}\frac{e}{s_{W}c_{W}}(s_{\alpha}c_{\beta}+\sqrt{\frac{3}{8}}c_{\alpha}s_{\beta}),\,g_{Z\eta H_{3}^{0}}=\mathfrak{i}\sqrt{\frac{2}{3}}\frac{e}{s_{W}c_{W}}(c_{\alpha}c_{\beta}-\sqrt{\frac{3}{8}}s_{\alpha}s_{\beta}),\,g_{ZH_{5}^{0}H_{3}^{0}}=-\mathfrak{i}\sqrt{\frac{1}{3}}\frac{e}{s_{W}c_{W}}c_{\beta},\nonumber \\
g_{ZZ\eta} & =\frac{e^{2}}{2s_{W}^{2}c_{W}^{2}}(s_{\alpha}c_{\beta}+\sqrt{\frac{8}{3}}c_{\alpha}s_{\beta}),\,g_{ZZh}=\frac{e^{2}}{2s_{W}^{2}c_{W}^{2}}(c_{\alpha}c_{\beta}-\sqrt{\frac{8}{3}}s_{\alpha}s_{\beta}),\,g_{ZH_{5}^{+}H_{3}^{-}}=\frac{e}{2s_{W}c_{W}}c_{\beta},\nonumber \\
g_{ZZH_{5}^{0}} & =-\frac{1}{\sqrt{3}}\frac{e^{2}}{s_{W}^{2}c_{W}^{2}}s_{\beta}\upsilon,\,g_{ZW^{+}H_{5}^{-}}=-\frac{e^{2}}{2s_{W}^{2}c_{W}}s_{\beta}\upsilon,\,g_{ZZh}^{SM}=\frac{e^{2}}{2s_{W}^{2}c_{W}^{2}}\upsilon.
\end{align}
Here, $g_{ZZh}^{SM}$ is the SM coupling. The couplings $g_{hXX,\eta XX}$
used in (\ref{eq:Rgg}), (\ref{eq:RgZ}) and (\ref{eq:eta-width})
are
\begin{align}
g_{hH_{5}^{++}H_{5}^{--}} & =g_{hH_{5}^{+}H_{5}^{-}}=-8\sqrt{3}(\lambda_{3}+\lambda_{4})\upsilon_{\xi}s_{\alpha}+(4\lambda_{2}+\lambda_{5})\upsilon_{\phi}c_{\alpha}-2\sqrt{3}\mu_{2}s_{\alpha},\nonumber \\
g_{hH_{3}^{+}H_{3}^{-}} & =-\frac{8}{\sqrt{3}}\Big(\frac{\sqrt{2}}{4}\lambda_{5}s_{\beta}c_{\beta}\upsilon_{\phi}+((\lambda_{3}+3\lambda_{4})\upsilon_{\xi}-\frac{3\mu_{2}}{4})c_{\beta}^{2}+\frac{3}{2}(\lambda_{2}+\frac{\lambda_{5}}{6})\upsilon_{\xi}+\frac{\mu_{1}}{24})s_{\beta}^{2}\Big)s_{\alpha}\nonumber \\
 & +2\sqrt{2}c_{\alpha}c_{\beta}s_{\beta}(\lambda_{5}\upsilon_{\xi}+\frac{\mu_{1}}{2})+4c_{\alpha}\Big(((\lambda_{2}-\frac{\lambda_{5}}{4})c_{\beta}^{2}+2\lambda_{1}s_{\beta}^{2})\upsilon_{\phi}\Big),\nonumber \\
g_{\eta hh} & =-2\sqrt{3}c_{\alpha}\Big(((\lambda_{5}-2\lambda_{2})\upsilon_{\xi}+\frac{\mu_{1}}{4})c_{\alpha}^{2}-4s_{\alpha}^{2}((\lambda_{3}+3\lambda_{4}+\frac{\lambda_{5}}{2}-\lambda_{2})\upsilon_{\xi}+\frac{\mu_{1}}{8}-\frac{\mu_{2}}{2})\Big)\nonumber \\
 & +4s_{\beta}\Big((\lambda_{5}+6\lambda_{1}-2\lambda_{2})c_{\alpha}^{2}-\frac{s_{\alpha}^{2}}{2}(\lambda_{5}-2\lambda_{2})\Big)\upsilon_{\phi},\nonumber \\
g_{\eta H_{5}^{++}H_{5}^{--}} & =g_{\eta H_{5}^{+}H_{5}^{-}}=g_{\eta H_{5}^{0}H_{5}^{0}}=8\sqrt{3}(\lambda_{3}+\lambda_{4})\upsilon_{\xi}s_{\alpha}+(4\lambda_{2}+\lambda_{5})\upsilon_{\phi}c_{\alpha}+2\sqrt{3}\mu_{2}s_{\alpha},\nonumber \\
g_{\eta H_{3}^{+}H_{3}^{-}} & =g_{\eta H_{3}^{0}H_{3}^{0}}=\frac{8}{\sqrt{3}}\Big(\frac{\sqrt{2}}{4}\lambda_{5}c_{\beta}s_{\beta}\upsilon_{\phi}+((\lambda_{3}+3\lambda_{4})\upsilon_{\xi}-\frac{3\mu_{2}}{4})c_{\beta}^{2}+\frac{3}{2}((\lambda_{2}+\frac{\lambda_{5}}{6})\upsilon_{\xi}+\frac{\mu_{1}}{24})s_{\beta}^{2}\Big)c_{\alpha}\nonumber \\
 & +2\sqrt{2}s_{\alpha}c_{\beta}s_{\beta}(\lambda_{5}\upsilon_{\xi}+\frac{\mu_{1}}{2})+4s_{\alpha}\Big((\lambda_{2}-\frac{\lambda_{5}}{4})c_{\beta}^{2}+2\lambda_{1}s^{2})\upsilon_{\phi}\Big),
\end{align}

The coefficients $C_{ZXX}$ used in (\ref{eq:RgZ}) are given by
\begin{align}
C_{ZH_{5}^{++}H_{5}^{--}} & =\frac{1-2s_{W}^{2}}{s_{W}c_{W}},~C_{ZH_{3}^{+}H_{3}^{-}}=C_{ZH_{5}^{+}H_{5}^{-}}=\frac{1-2s_{W}^{2}}{2s_{W}c_{W}}.
\end{align}

\section{WRONG MINIMA\label{app:WM}}

The GM scalar potential may have other minima than the EW one. It
is possible to get analytic formula for some these wrong minima, like
the ones below, but others require numerical efforts. The following
minima are possible only if the quantities inside the square-root
are positive.

In the $CP-even$ subspace: we have eight possible minima that corresponds
to $V_{i}^{0+}$,
\begin{equation}
\begin{array}{cc}
\{h_{\phi},\,h_{\chi},\,h_{\xi}\} & =\Big(\pm\frac{\sqrt{-\lambda_{1}m_{1}^{2}}}{2\lambda_{1}},0,0\Big),\,\Big(0,\pm\frac{\sqrt{-2m_{2}^{2}(2\lambda_{4}+\lambda_{3})}}{2(2\lambda_{4}+\lambda_{3})},0\Big),\\
 & \Big(0,0,\pm\frac{\sqrt{-m_{2}^{2}(\lambda_{4}+\lambda_{3})}}{2(\lambda_{4}+\lambda_{3})}\Big),\ \Big(0,\frac{1}{\lambda_{3}}\sqrt{\frac{-m_{2}^{2}\lambda_{3}^{2}-9\mu_{2}^{2}\lambda_{3}-9\mu_{2}^{2}\lambda_{4}}{2\lambda_{3}+4\lambda_{4}}},-\frac{3\mu_{2}}{2\lambda_{3}}\Big),\\
 & \Big(0,\,\pm\frac{\sqrt{3\mu_{2}\sqrt{-4m_{2}^{2}\lambda_{3}-12m_{2}^{2}\lambda_{4}+9\mu_{2}^{2}}-2m_{2}^{2}\lambda_{3}-6m_{2}^{2}\lambda_{4}+9\mu_{2}^{2}}}{2\lambda_{3}+6\lambda_{4}},\,\frac{3\mu_{2}+\sqrt{-4m_{2}^{2}\lambda_{3}-12m_{2}^{2}\lambda_{4}+9\mu_{2}^{2}}}{4\lambda_{3}+12\lambda_{4}}\Big),\\
 & \Big(0,\,\pm\frac{\sqrt{-3\mu_{2}\sqrt{-4m_{2}^{2}\lambda_{3}-12m_{2}^{2}\lambda_{4}+9\mu_{2}^{2}}-2m_{2}^{2}\lambda_{3}-6m_{2}^{2}\lambda_{4}+9\mu_{2}^{2}}}{2\lambda_{3}+6\lambda_{4}},\,\frac{3\mu_{2}+\sqrt{-4m_{2}^{2}\lambda_{3}-12m_{2}^{2}\lambda_{4}+9\mu_{2}^{2}}}{4\lambda_{3}+12\lambda_{4}}\Big),\\
 & \Big(0,\,\pm\frac{\sqrt{3\mu_{2}\sqrt{-4m_{2}^{2}\lambda_{3}-12m_{2}^{2}\lambda_{4}+9\mu_{2}^{2}}-2m_{2}^{2}\lambda_{3}-6m_{2}^{2}\lambda_{4}+9\mu_{2}^{2}}}{2\lambda_{3}+6\lambda_{4}},\,-\frac{-3\mu_{2}+\sqrt{-4m_{2}^{2}\lambda_{3}-12m_{2}^{2}\lambda_{4}+9\mu_{2}^{2}}}{4\lambda_{3}+12\lambda_{4}}\Big),\\
 & \Big(0,\,\pm\frac{\sqrt{-3\mu_{2}\sqrt{-4m_{2}^{2}\lambda_{3}-12m_{2}^{2}\lambda_{4}+9\mu_{2}^{2}}-2m_{2}^{2}\lambda_{3}-6m_{2}^{2}\lambda_{4}+9\mu_{2}^{2}}}{2\lambda_{3}+6\lambda_{4}},\,-\frac{-3\mu_{2}+\sqrt{-4m_{2}^{2}\lambda_{3}-12m_{2}^{2}\lambda_{4}+9\mu_{2}^{2}}}{4\lambda_{3}+12\lambda_{4}}\Big).
\end{array}\label{eq:C1}
\end{equation}

In the $CP-odd$ subspace: we got three possible minima that corresponds
to $V_{i}^{0-}$,
\begin{align}
\{a_{\phi},\,a_{\chi}\} & =\Big(\pm\frac{\sqrt{-\lambda_{1}m_{1}^{2}}}{2\lambda_{1}},0\Big),~\Big(0,\pm\frac{\sqrt{-2m_{2}^{2}(2\lambda_{4}+\lambda_{3})}}{2(2\lambda_{4}+\lambda_{3})}\Big),\nonumber \\
 & \Big(\sqrt{\frac{-8m_{1}^{2}\lambda_{3}-16m_{1}^{2}\lambda_{4}+8m_{2}^{2}\lambda_{2}-2m_{2}^{2}\lambda_{5}}{32\lambda_{1}\lambda_{3}+64\lambda_{1}\lambda_{4}-16\lambda_{2}^{2}+8\lambda_{2}\lambda_{5}-\lambda_{5}^{2}}},\,\sqrt{\frac{8m_{1}^{2}\lambda_{2}-2m_{1}^{2}\lambda_{5}-16m_{2}^{2}\lambda_{1}}{32\lambda_{1}\lambda_{3}+64\lambda_{1}\lambda_{4}-16\lambda_{2}^{2}+8\lambda_{2}\lambda_{5}-\lambda_{5}^{2}}}\Big).\label{eq:C2}
\end{align}

In the singly charged subspace: in this direction, we parametrized
the charged fields as $X^{\pm}=|X|e^{\pm i\varrho}$, and then we
found that the minima that correspond to $V_{i}^{\pm}$ do not depend
on the phases, i.e.,
\begin{equation}
\begin{array}{cc}
\{|\phi^{\pm}|,\,|\chi^{\pm}|,\,|\xi^{\pm}|\} & =\Big(\pm\frac{\sqrt{-2\lambda_{1}m_{1}^{2}}}{4\lambda_{1}},0,0\Big),\,\Big(0,\pm\frac{\sqrt{-2m_{2}^{2}(\lambda_{4}+\lambda_{3})}}{4(\lambda_{4}+\lambda_{3})},0\Big),\,\Big(0,0,\pm\frac{\sqrt{-2m_{2}^{2}(\lambda_{4}+\lambda_{3})}}{4(\lambda_{4}+\lambda_{3})}\Big),\\
 & \left(\sqrt{\frac{\lambda_{2}m_{1}^{2}-2\lambda_{1}m_{2}^{2}}{16\lambda_{1}\lambda_{3}+16\lambda_{1}\lambda_{4}-4\lambda_{2}^{2}}},\sqrt{\frac{-2\lambda_{3}m_{1}^{2}-2\lambda_{4}m_{1}^{2}+\lambda_{2}m_{2}^{2}}{16\lambda_{1}\lambda_{3}+16\lambda_{1}\lambda_{4}-4\lambda_{2}^{2}}},0\right),\left(\sqrt{\frac{-m_{2}^{2}}{16\lambda_{4}+8\lambda_{3}}},0,\sqrt{\frac{-m_{2}^{2}}{16\lambda_{4}+8\lambda_{3}}}\right),\\
 & \left(0,\sqrt{\frac{-2\lambda_{3}m_{1}^{2}-2\lambda_{4}m_{1}^{2}+\lambda_{2}m_{2}^{2}}{16\lambda_{1}\lambda_{3}+16\lambda_{1}\lambda_{4}-4\lambda_{2}^{2}}},\sqrt{\frac{\lambda_{2}m_{1}^{2}-2\lambda_{1}m_{2}^{2}}{16\lambda_{1}\lambda_{3}+16\lambda_{1}\lambda_{4}-4\lambda_{2}^{2}}}\right).
\end{array}\label{eq:C3}
\end{equation}

In the doubly charged subspace: in the doubly charged directions
we have only one possible minimum, which is given by
\begin{equation}
|\chi^{\pm\pm}|=\frac{\sqrt{-m_{2}^{2}(2\lambda_{4}+\lambda_{3})}}{2(2\lambda_{4}+\lambda_{3})}.\label{eq:C4}
\end{equation}

\textbf{Acknowledgements}: We would like to thank Abdesslam Arhrib
for his valuable comments. The work of A.A. is funded by the University
of Sharjah under the Research Projects No.
21021430107 ``\textit{Hunting for New Physics at Colliders}'' and No. 23021430135 ``\textit{Terascale Physics: Colliders vs Cosmology}''.

\end{document}